\documentclass[journal,onecolumn,12pt,draftclsnofoot]{IEEEtran}

\usepackage{subfigure}

\usepackage{cite}

\usepackage[dvipdfmx]{graphicx} 
\usepackage{bmpsize}

\usepackage[cmex10]{amsmath}
\usepackage{amsmath,amssymb,amsthm,amsfonts,latexsym,bbm,xspace,graphicx,float,mathtools,
verbatim} 
\usepackage{breqn}
\usepackage{algorithmic}

\usepackage{stackrel}

\usepackage[dvipsnames]{xcolor}

\definecolor{mypink1}{rgb}{0.858, 0.188, 0.478}
\definecolor{mypink2}{RGB}{219, 48, 122}
\definecolor{mypink3}{cmyk}{0, 0.7808, 0.4429, 0.1412}
\definecolor{mygray}{gray}{0.6}

\usepackage{array}

\usepackage[caption=false,font=normalsize,labelfont=sf,textfont=sf]{subfig}

\usepackage{algorithm}

\usepackage{amsthm}
\usepackage[utf8]{inputenc}
\usepackage{float}
\usepackage{cite}
\usepackage{pgfplots}
\usepackage[utf8]{inputenc}
\usepackage[T1]{fontenc}

\newtheorem{theorem}{Theorem}

\newtheorem{lemma}[theorem]{Lemma}

\newtheorem{cor}{Corollary}
\newtheorem{assumption}{Assumption}

\theoremstyle{definition}
\newtheorem{defn}{Definition}

\newcommand\numberthis{\addtocounter{equation}{1}\tag{\theequation}}

\def\BibTeX{{\rm B\kern-.05em{\sc i\kern-.025em b}\kern-.08em
    T\kern-.1667em\lower.7ex\hbox{E}\kern-.125emX}}

\begin{document}
%
\title{Private Weighted Random Walk Stochastic Gradient Descent}
%
%
%

        
\author{
	\IEEEauthorblockN{Ghadir Ayache, Salim El Rouayheb} \\
	\IEEEauthorblockA{
		Department of Electrical and Computer Engineering, Rutgers University \\
		Emails: \{ghadir.ayache,  salim.elrouayheb\}@rutgers.edu
	 }                    
}

\maketitle

\begin{abstract}
We consider a decentralized learning setting in which data is distributed over nodes in a graph. The goal is to learn a global model on the distributed data without involving any central entity that needs to be trusted. While gossip-based stochastic gradient descent (SGD) can be used to achieve this learning objective, it incurs high communication and computation costs.   
To speed up the convergence, we propose instead to study random walk based SGD in which a global model is updated based on a random walk on the graph. We propose two algorithms based on two types of random walks that achieve, in a decentralized way,  uniform sampling and importance sampling of the data. We provide a non-asymptotic analysis on the rate of convergence, taking into account the constants related to the data and the graph. Our numerical results show that the weighted random walk based algorithm has a better performance for high-variance data.   Moreover, we propose a privacy-preserving random walk algorithm that achieves local differential privacy based on a Gamma noise mechanism that we propose.   We also give numerical results on the convergence of this algorithm and show that it outperforms additive Laplace-based privacy mechanisms.


\end{abstract}


\IEEEpeerreviewmaketitle

\section{Introduction}
We consider the problem of designing a decentralized learning algorithm on data that is distributed among the nodes of a graph.   Each node in the graph has some local data, and we want to learn a model   by minimizing an empirical  loss function on   the collective data. 
We focus on applications such as  decentralized federated learning or    Internet-Of-Things (IoT) networks,  where the nodes, being phones or IoT devices,  have limited   communication and energy resources.  
A crucial constraint  that we impose is excluding the reliance on any central unit (parameter server, aggregator, etc.) that can communicate with all the nodes and orchestrate the learning algorithm.  Therefore,  
  we seek   decentralized algorithms that are  based only on local communication between neighboring nodes. Privacy is our main motivation for such algorithms since nodes do not  have to trust a central unit. Naturally, decentralized algorithms come with the added benefits of  avoiding single points of failure  and easily adapting to dynamic graphs. 
  
  Gossip-style gradient descent algorithms,  e.g.,\cite{Ram2009AsynchronousGA}, are a major contender for solving our problem.  In Gossip algorithms,  each node has a {\em local} model that is updated    and exchanged with the neighboring nodes in each iteration. Gossip algorithms can be efficient time-wise since nodes can update their model in parallel. However,  it incur  high communication and energy (battery) cost in order to wait for all the local models to  converge.    This diminishes the appeal of Gossip algorithms for the applications we have in mind. 
    Instead, we propose  to study decentralized algorithms based on random walks. A random walk passes  around a global model, and   in each iteration, activates   a node  which updates  the global model based on its local data. The activated node then passes the updated global model to a randomly chosen  neighbor. The random walk guarantees that every cost spent on the communication and computation resources goes to improve the global model.  Therefore, it outperforms Gossip algorithms in terms of these costs.   In this paper, we will study how the design of the random walk affects the convergence of the learning algorithm. Namely, we propose two algorithms that we call Uniform Random Walk SGD and Weighted Random Walk SGD and study their convergence.   Moreover, we consider the setting in which nodes do not completely trust their neighbors and  devise locally differential variants of these algorithms. 
    

\subsection{Previous work}

The two works that are most related to our work are \cite{MCGD} and \cite{Needell}.    The work in \cite{MCGD}  studies the random walk data sampling for stochastic gradient descent (SGD).  The work of \cite{Needell} focuses on importance sampling to speed up the convergence; however, it is implemented for a centralized   setting.


{A line of work on random walk based decentralized algorithm has been studied in the literature under the name of incremental methods. Early works on incremental methods by \cite{Johansson2007ASP,Johansson2009ARI,Nedic2009DistributedSM}  have established theoretical convergence guarantees for different convex problem settings and using first-order methods of stochastic gradient updates. In \cite{Wai2018SUCAGSU}, more advanced stochastic updates is been employed, namely CIAG \cite{Wai2017CurvatureaidedIA}, to improve the convergence guarantees for strongly convex objective by implementing unbiased total gradient tracking technique that uses Hessian information to accelerate convergence; they show that their algorithm converges linearly with high probability.} The work of  \cite{MCGD} proposed   to speed-up the convergence by using non-reversible random walks. In \cite{Mao2020WalkmanAC}, the authors studied the convergence analysis of random walk learning algorithm for  the alternating direction method of multipliers (ADMM).    

Gossip-based algorithms   are also popular approach that has been widely studied, with application to averaging  \cite{Boyd2006RandomizedGA} and  various stochastic gradient learning   \cite{Nedic2009DistributedSM,Koloskova2019DecentralizedSO,Aysal2009BroadcastGA,Duchi2012DualAF} .

In terms of privacy, many works study differential privacy mechanisms for SGD algorithms. One approach for privacy is local differential privacy (LDP) \cite{duchi2013local,kairouz2014extremal}, where privacy is applied locally, which is well-suited for decentralized algorithms. Privacy mechanisms for SGD suggest protecting the output of every iteration by output perturbation
 \cite{Wu2016DifferentiallyPS} or by gradient perturbation and  quantization (e.g.,\cite{Bassily2014PrivateER,Gandikota2019vqSGDVQ,Xiong2016RandomizedRW, Chen2020BreakingTC}).




\subsection{Contributions}

In our work, we are interested in   decentralized settings where the nodes have limited computation and communication power. Moreover, we assume that the nodes do not fully trust each others with their data and require privacy guarantees. For these reasons, we focus on first-order methods where the information exchanged between neighboring nodes is restricted to the model at a given iteration, serving both our communication and privacy constraints. Moreover,  the update step  involves only the gradient of the local loss function but no higher order derivatives, limiting the computation cost at the nodes.
In this constrained setting, we study the design the random walk to speed up the convergence by simulating importance sampling.

It is known that the natural random walk on the vertices of a graph that picks one of the neighbors uniformly at random, will end up visiting  nodes proportionally to their degree. The consequence of this is that the underlying SGD learning algorithm will be effectively sampling with higher probability  the data on highly connected nodes. This sampling, which is biased by the graph topology,  may or may not be a good choice for  speeding-up the convergence of the algorithm, depending on which nodes have the important  data. 

 We propose two alternate Markov-Hasting sampling schemes to address this bias, and study the convergence of the resulting algorithms: 
 (i) {\em Uniform Random Walk SGD (Algorithm~\ref{Alg1})}, in which  all the nodes   are visited  equally likely in the stationary regime   irrespective of the node degrees.  This emulates uniform data sampling in the centralized case ;
 (ii) {\em The Weighted Random Walk SGD (Algorithm~\ref{Alg2})}, in which  the nodes are sampled proportional to the gradient-Lipschitz constant of their local loss function, rather  than their degree. This emulates importance sampling \cite{ Needell , Zhao2014StochasticOW} in the centralized case. 

We study the rate of convergence of both algorithms. The asymptotic rate of convergence of both algorithms is same as that of the natural random walk, namely, $\mathcal{O}(\frac{1}{k^{ 1-q}})$, where $k$ is the number of iterations and $q \in (\frac{1}{2},\,1)$.  However, the constants are different and depend  on the gradient-Lipschitz constants of the local loss functions and the graph spectral properties. Our numerical simulations highlight how these constants affect the non-asymptotic behavior of these algorithms. we observe that in high variance data regime   Algorithm~\ref{Alg2} outperforms Algorithm~\ref{Alg1}, and  the opposite happens in a low variance data regime.   These observations are in conformity with the behavior of importance sampling in \cite{Needell} for the centralized case. This is expected   since the random walks were designed to achieve, in the stationary regime, the same sampling strategy of the centralized case. The challenge, however, is with the proof techniques, which build on  the random walk learning results in  \cite {MCGD} and    importance sampling in  \cite{Needell}.  

In addition, we propose a third algorithm,      Private Weighted Random Walk SGD (Algorithm~\ref{Alg3}), that ensures local differential privacy of the local data.  
  We propose a privacy mechanism that uses   Gamma noise and characterize the tradeoff between the noise level and the privacy level. We also give numerical results on the convergence of  Algorithm~\ref{Alg3} and show that it outperforms additive Laplace-based privacy mechanisms.

Parts of these results pertaining to Algorithms~\ref{Alg1} and \ref{Alg2} have already appeared in a conference paper \cite{Ayache2019RandomWG}.


\subsection{Organization}
The rest of the paper is organized as follows.  We formulate the problem  in  Section~\ref{sectionII}.
We present Algorithm~\ref{Alg1} and Algorithm~\ref{Alg2}  in Section~\ref{sectionIII}. We present the convergence result of both algorithms in Subsection~\ref{subesctionC}. In Section~\ref{sectionIV}, we define the Gamma mechanism and propose a  privacy-preserving version of Algorithm~\ref{Alg2}. Finally, we present numerical results of our algorithms in Section~\ref{sectionV}. We conclude in Section~\ref{sectionIV}.  The proofs of the technical results are deferred to the appendices.

\section{{Problem Description}}
\label{sectionII}
\subsection{Problem Setting}

\noindent \textbf{Network Model:} 
We consider a communication network represented by an undirected  graph $\mathcal{G}(V,\,E)$ with $V =\left[N\right] := \{1, \ldots, N\}$ is the set of nodes and $E \subseteq V \times V$ is the set of edges. 
Two nodes   in $\mathcal{G}$   connected by an edge in $E$ can communicate and exchange information. We refer to such two nodes as neighbors.  We denote the set of neighbors of a node $i\in V$ by $\mathcal{N}(i)$, and   its degree  by  $\deg(i) := |\mathcal{N}(i)|$.
We assume that the graph $\mathcal{G}$ is connected, i.e., there exits a path between any two nodes in $V$. Moreover, we assume the presence of a self-loop at every node, i.e, $\{(i,\,i)\,:\,i=1,\,2,\, ...,\, N\} \subset E.$ We assume that the set $\mathcal{N}(i)$ does not contain $i$. 

\noindent \textbf{Data and Learning Objective:} Each node $i\in [N]$ has a feature vector $x_i\in \mathcal{X}\subseteq \mathbb{R}^d$,  where  $\mathcal{X}$ is the feature space and $d$ is the   number of features. Moreover, node $i$ has a label $y_i\in \mathbb{R}$ corresponding to $x_i$.  

The objective is to learn a prediction function on the collective data distributed over all the nodes by learning a global model $w \in \mathcal{W}$, where  $\mathcal{W}$ is a closed and bounded subset in $\mathbb{R}^d$. Thus, an optimal model $w^*$ solves
\begin{align}
\underset{w\in\mathcal{W}}{\text{minimize}} \frac{1}{N}\stackrel[i=1]{N}{\sum}&f_i\left(w;\,x_i,\,y_i\right), \label{eq:1} 
\end{align}
where $f_i\left(.\right)$ is the local loss function at node $i$.  We denote the global loss function on all the data by \begin{align*}
    f\left(w\right) = \frac{1}{N}\stackrel[i=1]{N}{\sum}f_i\left(w;\,x_i,\,y_i\right).
\end{align*} We will assume convexity and gradient-Lipschitzness of the local loss functions as described in Assumption~\ref{assumption1}.
  \begin{assumption}
  \label{assumption1}
The local loss function $f_i$ for each node $i \in V$ is a convex function on $\mathcal{W} \in \mathbb{R}^d$ and has an $L_i$-Lipschitz continuous gradient; that is, for any $w$,\,$w' \in \mathcal{W}$, there exists a constant $L_i>0 $ such that
\begin{align*}
  \left\Vert \nabla f_i\left(w\right)-\nabla f_i\left(w'\right)\right\Vert _{2}\leq L_{i}\left\Vert w-w'\right\Vert_2.  
\end{align*}\end{assumption} 
To solve the optimization problem in \eqref{eq:1}, we perform an iterative stochastic gradient descent (SGD) method. At   each iteration $k$, one random node $i$ in the network will perform a gradient descent update and project the result back into the feasible set $\mathcal{W}$ as follows:
\begin{align}
\label{eq:2}
 w^{(k+1)}= {\bf{\Pi}}_{\mathcal{W}}\left(w^{(k)}-\gamma^{(k)}\hat{\nabla} f\left(w^{(k)}\right)\right),
 \end{align} 
  where $\gamma^{(k)}$ is the step size, $w^{(k)}$ is the global model update at iteration $k$, and $\hat{\nabla} f\left(w^{(k)}\right)$ is a gradient estimate of the global loss function $f(.)$ based on one node $i$ local data. For convergence guarantees \cite{Boyd2007SubgradientM}, we use a decreasing step size that satisfies 
 \begin{align}
     \stackrel[k=1]{\infty}{\sum}\gamma^{(k)}=+\infty, \,\,\,\,\,\,\, \text{and} \,\,\,\,\,\,\, \stackrel[k=1]{\infty}{\sum}\ln k.{(\gamma^{(k)})}^{2}<+\infty.
     \label{StepsizeEq}
 \end{align}
 In particular, we will take $\gamma^{(k)}=\frac{1}{k^q}\,$ for $\,0.5<q<1$ which satisfy the conditions in \eqref{StepsizeEq}.
 
Our goal is to implement a decentralized algorithm based on a Markov random walk sampling  decision on the graph $\mathcal{G}$
to solve the learning problem in \eqref{eq:1} with privacy guarantees, as explained in the next two sections.

\subsection{Random Walk Learning Algorithm}


We seek a decentralized algorithm that will learn the model $w^*$ without involving a central entity, and  in which nodes exchange information only with their neighbors.
Towards that goal, we   design a random walk on $\mathcal{G}$   that activates the node that it visits to update the global model based on the local data of the activated node. 
The algorithm is initiated at one node picked uniformly at random to be activated at the first iteration to perform the update in \eqref{eq:2}. Afterwards, at every iteration $k = 1,\,2,\,...\,$,   the global model iterate $w^{(k)}$ is passed to a randomly chosen node $i^{(k)} \in V$ to update it using its local data.  Given the connection constraint, at iteration $k$, the activated node $i^{(k)}$ is a neighbor node of the previously activated node $i^{(k-1)}$, i.e., $i^{(k)} \in \mathcal{N}(i^{(k-1)})$.
The process above defines a random walk on $\mathcal{G}$ that we model by a Markov chain.
The sequence of active nodes $i^{(k)}$ are the states of the Markov chain with state-space $ V $ and transition matrix $ P $ that inherits the same connection structure of the underlying connected graph $\mathcal{G}$. For convergence guarantees, we will make  the next assumption:  

\begin{assumption}

\label{assumption2}
The Markov chain $(i^{(k)})_{k \in \mathbb{N}}$ defined on the finite state space $V$ with  homogeneous transition matrix $P$ is irreducible and aperiodic and has a stationary distribution $\pi$ on $V$. 
\end{assumption}


{The problem is to design the transition probabilities in $ P $ to speed up the convergence of the decentralized algorithm by only using   the local information available to the nodes, namely their degrees and  Lipschitz constants.  }


\subsection{Local Differential Privacy}
In addition to designing a decentralized random walk learning algorithm that speeds up convergence, we aim for a privacy-preserving algorithm to protect each node data from being revealed by any other node including its neighbors.



We adopt local differential privacy (LDP) \cite{duchi2013local,kairouz2014extremal} as a privacy measure that is  well-suited to our   decentralized setting, since we excluded the involvement of any central aggregator that would coordinate the learning process. Accordingly, nodes will share a noisy version of the messages that needs to be exchanged with the neighbors in a way that ensures a desired level of privacy for the nodes' local data.

Consider a message $M(x_i)$ that is to be sent from node $i$ to one of its neighbors.   $M$ is a function of the node's local data   $M: \mathcal{X} \rightarrow \mathbb{M}$, where $\mathbb{M}$ is the image space. Let $\mathcal{R}: \mathbb{M} \rightarrow Img \mathcal{R}$ be the privatization scheme that   node $i$ will perform on the message to share. For any two possible data points $x_i,\, x'_i \in \mathcal{X}$ that could be owned by a node $i$,  the  $(\epsilon,\,\delta)$-LDP defined as follows:
\begin{defn}[Local Differential Privacy \cite{duchi2013local}]
\label{defDP}
A randomization mechanism $\mathcal{R}$ is $(\epsilon,\,\delta)$-LDP, if for any $x_i,\, x'_i \in \mathcal{X}$ and for any $r \in  {Img} \mathcal{R} $, we have\begin{align}P\left(\mathcal{R}\left(M\left(x_{i}\right)\right)=r\right)\leq e^{\epsilon}P\left(\mathcal{R}\left(M\left(x'_{i}\right)\right)=r\right)+\delta,\label{eq:4}\end{align}
\end{defn}\noindent where $\epsilon \geq 0$ quantifies the privacy level, and  $\delta \in [0,1]$ quantifies the allowed probability of violating the privacy bound \cite{Dwork2014TheAF,balle }.






A drawback of 
  LDP noising mechanisms is that it    will affect the utility of the exchanged messages and, as a result, slow the convergence of decentralized learning algorithm. We propose an LDP mechanism based on Gamma noise  that achieves an attractive tradeoff between privacy and convergence, and outperforms generic additive-noise (Gaussian or Laplace) LDP mechanisms.

For ease of reference, we summarize our notation in Table I.

\begin{table}[t]
\centering
\begin{tabular}{c|l|}
 
\hline
\multicolumn{1}{|l|} {$\mathcal{G}$}	 & 	Graph representing the network\\ \hline
\multicolumn{1}{|l|}{$N$}	 & Total number of nodes in $\mathcal{G}$ \\ \hline 
\multicolumn{1}{|l|} {$V$}	 & 	Set of nodes in {$\mathcal{G}$}\\ \hline
\multicolumn{1}{|l|} {$E$}	 & 	Set of edges in {$\mathcal{G}$}\\ \hline
\multicolumn{1}{|l|} {$\deg\left(i\right)$}	 & 	Degree of node $i$\\ \hline
\multicolumn{1}{|l|} {$x_{i}$}	 &  Data feature vector of node $i$\\ \hline
\multicolumn{1}{|l|} {$y_{i}$}	 &  Data label of node $i$ \\ \hline
\multicolumn{1}{|l|} {$f_{i}$}	 & 	Local loss function of node $i $\\ \hline
\multicolumn{1}{|l|} {$L_{i}$}	 &  Gradient-Lipschitz constant of $f_{i}$ \\ \hline
\multicolumn{1}{|l|} {$f$}	 & 	Global loss function \\ \hline
\multicolumn{1}{|l|} {$w$}	 & 	Data model \\ \hline
\multicolumn{1}{|l|} {$w^{*}$}	 & 	Data optimal model \\ \hline
\multicolumn{1}{|l|} {$w^{\left(k\right)}$}	 & 	Model update at iteration $k$ \\ \hline
\multicolumn{1}{|l|} {$\sup L$}	 & 	Maximum gradient-Lipschitz constant \\ \hline
\multicolumn{1}{|l|} {$\inf L$}	 & 	Minimum gradient-Lipschitz constant \\ \hline
\multicolumn{1}{|l|} {$\bar{L}$}	 & 	Average  gradient-Lipschitz constant: $\bar{L} = \frac{L_{1}+L_{2}+...+L_{N}}{N}$ \\ \hline
\multicolumn{1}{|l|} {$i^{\left(k\right)}$}	 & 	Node visited at time $k$ \\ \hline
\multicolumn{1}{|l|} {$\epsilon$}	 & Privacy level \\ \hline
\multicolumn{1}{|l|} {$1-\delta$}	 & 	Privacy confidence level \\ \hline
\multicolumn{1}{|l|} {$\pi_{u}$}	 & 	Stationary distribution for uniform random walk \\ \hline
\multicolumn{1}{|l|} {$P_{u}$}	 & 	Transition matrix for uniform random walk \\ \hline
\multicolumn{1}{|l|} {$\pi_{w}$}	 & 	Stationary distribution for weighted random walk \\ \hline
\multicolumn{1}{|l|} {$P_{w}$}	 & 	Transition matrix for weighted random walk \\ \hline
\multicolumn{1}{|l|} {$\mathcal{R}\left(L_{i}\right)	$}	 & Noisy version of $L_{i}$, output of the privacy mechanism \\ \hline
\multicolumn{1}{|l|} {$\pi_{w,\,\mathcal{R}}$}	 & 	Stationary distribution for noisy (private) weighted random walk \\ \hline

\end{tabular} \label{Notation}
\caption{Notation}
\end{table}





\section{{Decentralized Weighted Random Walk SGD}}
\label{sectionIII}
We focus in this section on the design of   the random walk learning algorithm without the privacy constraint. 
To decide on the transition probabilities of the random walk, a natural choice would be to pick the next node   uniformly at random from the neighbors of the current active node. {This gives a stationary distribution $\pi$ proportional to the degree of each node, i.e, $\pi(i)\sim \deg (i)$.  Therefore, Assumption \ref{assumption2} implies   the fraction of time a node is activated is directly proportional to its degree \cite{Levin} . The effect of this random walk on the learning algorithm is that, in the update step of \eqref{eq:2}  the data is sampled proportional to the degree of the node that is carrying it.
 This node degree biased sampling may not be a favorable choice for speeding up the convergence of our decentralized learning algorithm.}
To counterbalance this bias, we propose two alternate sampling schemes and study the convergence of the resulting algorithms.
\begin{enumerate}

\item    {\bf Uniform Random Walk SGD}, in which  all the nodes   are visited  equally likely in the stationary regime   irrespective of their degree. This guarantees that the data points are eventually sampled uniformly at random irrespective of the graph topology.
\item {\bf   Weighted Random Walk SGD}, in which the nodes are sampled proportional to the gradient-Lipschitz constant of their local loss function, rather  than their degree. The motivation is to achieve importance sampling \cite{ Needell , Zhao2014StochasticOW} in a decentralized fashion. 
\end{enumerate}


\subsection{Uniform Random Walk SGD} 
The  Uniform Random Walk SGD algorithm aims at sampling the data points held by the graph nodes uniformly at random, similarly to what is done 
in standard centralized SGD \cite{Robbins2007ASA, Bottou2010LargeScaleML,Bach2011NonAsymptoticAO }. We achieve the  uniform data sampling  by designing a random walk with a  transition matrix $P_u$  using only local node information, such that    the chain converges to the stationary distribution  $\pi_{
u}$ such that\begin{align*}
\ensuremath{\pi_u\left(i\right)=\frac{1}{N},\,\,\,\,\,\ \forall i \in V.}
\end{align*} 
We implement the  Metropolis Hasting (MH) decision rule to design the transition probabilities, so the random walk converges to the desired stationary $ \pi_u $ \cite{Levin,lee,MCGD}.
The MH rule can be described as follows:
\begin{enumerate}
\item At the  $k^{th}$ step of the random walk, the active node $i^{(k)}$ selects uniformly at random one of its neighbors, say $j$, as a candidate to be the next active node.  This selection gets accepted with probability
\begin{align*}a_u\left(i^{(k)},\,j\right)=\min\left(1,\,\frac{\deg\left(i^{(k)}\right)}{\deg\left(j\right)}\right).\end{align*}
Upon the acceptance, we have $i^{(k+1)}=j$.
\item Otherwise, if the candidate node gets rejected,  the random walk stays at the same node, i.e.,  $i^{(k+1)} = i^{(k)}$.
 \end{enumerate}

Therefore, the  transition matrix $P_u$ is given by

\begin{align*}
    P_u\left(i,\,j\right)=
    \begin{cases}
\frac{1}{\deg\left(i\right)}\min\,\left\{ 1,\,\frac{\deg\left(i\right)}{\deg\left(j\right)}\right\}  & j\neq i\:\text{ and }\,j\in\mathcal{N}\left(i\right),\\
1-\sum_{j\in\mathcal{N}\left(i\right)}\frac{1}{\deg\left(i\right)}\min\,\left\{ 1,\,\frac{\deg\left(i\right)}{\deg\left(j\right)}\right\}  & j=i,\, \text{ and }\\
0 & \text{ otherwise. }
\end{cases}
\end{align*}

This  random walk does conform with Assumption~\ref{assumption2} and it uses local information only.
By  the Ergodic theorem \cite{Levin} the above random walk that converges to a uniform stationary distribution, samples all the states (nodes) uniformly at random on the long-term run. 

The Uniform Random Walk SGD implements the uniform random walk through the MH rule above. And, once a node is activated it updates the global model based on \eqref{eq:2}.

\begin{algorithm}[htb]
   \caption{Uniform Random Walk SGD}
\begin{algorithmic}
  \STATE{ {\bfseries Initialization:} Initial node $i^{(1)}$ chosen uniformly at random from $[N]$, Initial model $w^{(1)}$ chosen uniformly at random from $\mathcal{W}$}

   \FOR{$k=1$ {\bfseries to} $T$}
   
   \STATE $ \hat{\nabla} f\left(w^{(k)}\right) = {\nabla} f_i\left(w^{(k)}\right)$.
    \STATE $w^{(k+1)}={\bf{\Pi}}_{\mathcal{W}}\left(w^{(k)}-\gamma^{(k)}\hat{\nabla} f\left(w^{(k)}\right)\right)$  
   \STATE Choose node $j$ uniformly at random from $\mathcal{N}\left(i^{(k)}\right).$
   \STATE Generate $p \sim U\left(0,\,1\right)$ where $U$ is the uniform distribution.
   \IF{$p\leq \min \, \left\{ 1,\, \frac{\deg\left(i^{(k)}\right)}{\deg\left(j\right)} \right\}$} 
   \STATE $i^{(k+1)}\leftarrow j $
   \ELSE 
   \STATE $i^{(k+1)}\leftarrow i^{(k)}$
   \ENDIF
   
   \ENDFOR
   \STATE {\bfseries Return:} $w^{(T)}.$  
\end{algorithmic}
\label{Alg1}
\end{algorithm}
In subsection III-C, we give the   convergence analysis for this algorithm.
\subsection{Weighted Random Walk SGD}

To speed up the  convergence, we propose a decentralized sampling that mimics centralized  importance sampling, which    consists of selecting more often the more informative data points \cite{Needell,Zhao2014StochasticOW,Chen2019FastAA, Borsos2019OnlineVR,Alain2015VarianceRI, bouchard2015online}.  In our our analysis, we will take  the data importance to be reflected through the gradient-Lipschitz constant of the node's local loss function  \cite{Needell}.  Again, we   utilize the MH decision rule for the random walk to achieve a stationary distribution proportional to the node gradient-Lipschitz constant as follows
\begin{align*}
\pi_w \left(i \right) = \frac{L_i}{\sum_{j=1}^N L_j}\,\,\,\,\,\ \forall i \in V.
\end{align*}
The random walk proceeds as previously explained, except for the probability of accepting the candidate node, which is now

\begin{align}a_{w}\left(i^{(k)},\,j\right)=\min\left(1,\,\frac{L_{j}}{L_{i^{(k)}}}\frac{\deg\left(i^{(k)}\right)}{\deg\left(j\right)}\right).\label{Metropolis}\end{align}

Consequently, the new transition matrix $P_w$ generating the weighted random walk is given by
\begin{align*}\label{eq:wrw}
 P_w \left(i,\,j\right) =
  \begin{cases}
\frac{1}{\deg\left(i\right)}\min\,\left\{ 1,\,\frac{L_{j}}{L_{i}}\frac{\deg\left(i\right)}{\deg\left(j\right)}\right\}  &  \hspace{-0.8cm} j\neq i\:\text{and}\,j\in\mathcal{N}\left(i\right),\\
1-\sum_{j\in\mathcal{N}\left(i\right)}\frac{1}{\deg\left(i\right)}\min\,\left\{ 1,\,\frac{L_{j}}{L_{i}}\frac{\deg\left(i\right)}{\deg\left(j\right)}\right\}  & j=i,\, \text{and}\\
0 & \text{otherwise}.
\end{cases}
\end{align*}


\begin{algorithm}[htb]
   \caption{Weighted Random Walk SGD}
\begin{algorithmic}
   \STATE{ {\bfseries Initialization:} Initial node $i^{(1)}$ chosen uniformly at random from $[N]$, Initial model $w^{(1)}$ chosen uniformly at random from $\mathcal{W}$}
   
      \STATE {Every node $i$ shares $L_i$ and $\deg(i)$ it with its neighbors.}

   \FOR{$k=1$ {\bfseries to} $T$}
    
   \STATE $ \hat{\nabla} f\left(w^{(k)}\right) = {\frac{\bar{L}}{L_{i^{(k)}}}}{\nabla}f_{i^{(k)}}\left(w^{(k)}\right)$
    \STATE $w^{(k+1)}={\bf{\Pi}}_{\mathcal{W}}\left(w^{(k)}-{\gamma^{(k)}}\hat{\nabla} f\left(w^{(k)}\right)\right)$
   \STATE Choose node $j$ uniformly at random from $\mathcal{N}\left(i^{(k)}\right).$ 
   \STATE Generate $p \sim U\left(0,\,1\right)$ where $U$ is the uniform distribution.
   \IF{$p\leq \min \, \left\{1,\,\frac{L_j}{L_{i^{(k)}}}\frac{\deg\left(i^{(k)}\right)}{\deg\left(j\right)}\right\}$}
   \STATE $i^{(k+1)}\leftarrow j $
   \ELSE 
   \STATE $i^{(k+1)}\leftarrow i^{(k)}$
   \ENDIF
   \ENDFOR
   \STATE {\bfseries Return:} $w^{(T)}.$    
\end{algorithmic}
\label{Alg2}
\end{algorithm}

\subsection{Convergence Analysis}
\label{subesctionC}

We summarize here our main results on the convergence of  Algorithms~\ref{Alg1} and~\ref{Alg2}. Our aim is to characterize how the design of the random walk affects the non-asymptotic rate of convergence of the algorithms. To give a bit of context, for convex and gradient-Lispchitz objective functions, 
 centralized SGD    has an asymptotic convergence rate in the  order of $\mathcal{O}(\frac{1}{\sqrt{k}})$ (after $k$ iterations)  for  diminishing step-size  and for independent data sampling \cite{Bottou2007TheTO, Bach2013NonstronglyconvexSS}. For our decentralized setting, in which data sampling is not independent and is constrained by the graph topology, random walk SGD algorithms can approach the same rate of convergence  for convex global loss function   \cite{Duchi2011ErgodicMD,MCGD,Johansson2007ASP,Johansson2009ARI,Ram2009AsynchronousGA}. Namely, for    a step-size $\gamma^{(k)} = \frac{1}{k^{q}}$ with $\frac{1}{2},  <q<1$, the work in \cite{MCGD}
  proves a rate  of convergence  $\mathcal{O}(\frac{1}{k^{1-q}})$.  
  
  It was shown in \cite{Duchi2011ErgodicMD} that $\Omega (\frac{1}{\sqrt{k}})$,   $k$ being the number of SGD updates, is a fundamental lower bound on the convergence rate  for convex optimization within the stochastic first-order oracle model with no access to independent sampling.
  
  Our proposed algorithms will also   have this rate of convergence. However, the choice of the random walk will affect the constants in the rate of convergence and can lead to a speed-up in the non-asymptotic regime. Theorems~\ref{tUniform} and ~\ref  {tWeighted}  characterize these constants of our proposed algorithms.

We  denote the eigenvalues of any the transition matrix $P$ of the random walk by
$\lambda_{1}=1>\lambda_{2}\geq...\geq\lambda_{N}$, and define 
$\lambda_{P}=\frac{\underset{}{\max}{\left\{ \left|\lambda_{2}\right|,\left|\,\lambda_{N}\right|\right\} }+1}{2}.$

\begin{theorem}[Convergence of   Algorithm~\ref{Alg1}]
\label{tUniform}
Under Assumptions~\ref{assumption1} and ~\ref{assumption2},   the Uniform Random Walk SGD  (Algorithm~\ref{Alg1}) converges in the mean sense, i.e.,  
\begin{align}\underset{k\rightarrow \infty}{\lim\,}\mathbb{E}\left(f\left(w^{(k)}\right)-f\left(w^{*}\right)\right)=0.\end{align} 
Moreover, its rate of convergence, for a step size $\gamma^{(k)}=\frac{1}{k^q}$,   $0.5<q<1$, is
\begin{align*} & \mathbb{E}\left[f\left(\bar{w}^{(k)}\right)-f\left(w^{*}\right)\right] =O\left(\frac{\max\left\{\sigma^{2},\,\left(\sup  L\right)^{2},\,\frac{1}{\ln\left(1/\lambda_{P_u}\right)}\right\}}{k^{1-q}}\right),\numberthis\label{boundUniform}
\end{align*}
 
where, 
  $ \bar{w}^{(k)} =\frac{1}{\stackrel[n=1]{k}{\sum}\gamma^{(n)}} \stackrel[m=1] {k}{\sum} \gamma^{(m)}w^{(m)}$, 
 $ \sup L = \underset{i\in V}{\max} L_i$ and    the residual  $\sigma^{2}=\underset{w^{*}}{\max}\,\mathbb{E}\left[\left\Vert \nabla f_{i}\left(w^{*}\right)\right\Vert _{2}^{2}\right]$.

\end{theorem}
Note the technicality that the convergence is in terms of $ \bar{w}^{(k)}$, instead of  $w ^{(k)}$, which is weighted average of the $w ^{(k)}$s. This is due to the proof technique.   
Next, we give the bound on the convergence rate for the Weighted Random Walk SGD.


\begin{theorem}[Convergence rate of   Algorithm~\ref{Alg2}]
\label{tWeighted}

Under assumptions~\ref{assumption1} and ~\ref{assumption2},   the Weighted Random Walk SGD  (Algorithm~\ref{Alg2}) converges in the mean sense, i.e., 
\begin{align}\underset{k\rightarrow \infty}{\lim\,}\mathbb{E}\left(f\left(w^{(k)}\right)-f\left(w^{*}\right)\right)=0.\end{align}

Moreover, its   rate of convergence,     for a step size $\gamma^{(k)}=\frac{1}{k^q}$ when $0.5<q<1$, is

\begin{align}\mathbb{E}\left[f\left(\bar{w}^{(k)}\right)-f\left(w^{*}\right)\right]=O\left(\frac{\max\left\{ \frac{\bar{L}}{\inf L}\sigma^{2},\,\left(\bar{{L}}\right)^{2},\,\frac{1}{\ln(1/\lambda_{P_w})}\right\} }{k^{1-q}}\right),
\label{boundWeighted}
\end{align}
where $ \inf L = \underset{i\in V}{\min} \, L_i$ and $\bar{L} = \frac{\sum_i L_i}{N}$.
\end{theorem}

The results on the rate of convergence stated in the above theorems  show the tradeoff  that the two random walks offer. The weighted sampling for random walk SGD improves the bound on convergence from being a function of     $\sup L$ to be a function of the  average $\bar{L}$. However,    it amplifies the   effect of the residual represented by $\sigma^2$. 
The numerical results will show later that in high variance data regime resulting in a wide range for the values of the Lipschitz constants, the improvement brought by $\bar{L}$ dominates over the residual amplification, so Algorithm~\ref{Alg2} outperforms Algorithm~\ref{Alg1}. However, the opposite happens for a low variance data regime where the data variance is low, and therefore the gap between $\sup L$ and  $\bar{L}$ is not significant. Therefore, these observations are in conformity with the behavior of importance sampling in \cite{Needell} for the centralized case. 
The proof of Theorem~\ref{tUniform} follows the same steps as in   \cite {MCGD} with  incorporation  of the Lipschitzness assumption and we provide it for completeness and for comparison with  Theorem~\ref{tWeighted}.      The detailed  proofs for both theorems can be found in Appendices \ref{Appendix1} and \ref{AppendixProofWeighted}.

\subsection{Comparison to Gossip Algorithm:}

To better understand the performance of our two algorithms, we compare them to Gossip algorithm.
 One of the most common approaches of the Gossip algorithms is the synchronous version  \cite{Nedic2009DistributedSM, Shi2015EXTRAAE, Chen2012FastDF,Jakoveti2014FastDG,Chen2019FastAA}, in which at every round, all nodes exchange and update their local models. For convex problems, the convergence rate is $O(1/\sqrt{k})$ \cite{Koloskova2020AUT},  where $k$ is the algorithm round in which $|\mathcal{E}|$ communication links are activated and $O(N)$ computations are performed. Therefore, the convergence rate as function of communication cost is  $O({|\mathcal{E}|}/{\sqrt{k}})$,  and as a function of computation cost is $O({N}/{\sqrt{k}})$. However, the convergence rate of Algorithms~ 1 and~2  as a function of  computation and communication cost  approaches a rate of $O(1/\sqrt{k})$. That is because at each round at most\footnote{ The upper bound on the number of activated links stems from  the fact that in  Algorithms~1 and~2   a node will probabilistically choose to  perform multiple  updates on its data in consecutive rounds without exchanging information with its neighbours, making these rounds communication-free.   } one communication link is activated  and one SGD update is performed.

\section{Differential private Random Walk SGD}
\label{sectionIV}

In this section, we propose an LDP mechanism to be applied to   the messages exchanged within   Algorithm~\ref{Alg2} to make it privacy-preserving.    
 Algorithm~\ref{Alg2} requires sharing two pieces of information between two neighbouring nodes:

\begin{enumerate}
    \item The gradient-Lipschitz constants: Algorithm~\ref{Alg2} requires every node $i\in V$ to share the gradient-Lipschitz constant $L_i$  with all its neighbors to be able to implement the MH rule of \eqref{Metropolis}. This can leak important information about the local data at the nodes.   
   For instance, for    linear regression loss function, we have $L_i = N \Vert x_i \Vert ^2_2$. 
    \item The model updates: An activated node needs to receive the latest updated version of the model $w^{(k)}$, which naturally contains information about the data on the nodes visited so far.  
\end{enumerate}

 The literature on privacy-preserving SGD has extensively studied the problem of designing LDP mechanisms for the model update or the gradient through model perturbation
 (e.g., \cite{Wu2016DifferentiallyPS}) or gradient perturbation and  quantization (e.g.,\cite{Bassily2014PrivateER,Gandikota2019vqSGDVQ,Xiong2016RandomizedRW, Chen2020BreakingTC}).  
 Within that space of works, the new aspect of our problem is the sharing of the gradient-Lipschitz  constants. For this reason, we focus in our analysis on    characterizing how sharing privatized (noisy) versions of these constants would  affect the    random walk  learning algorithm, and assume that the models are shared in the clear\footnote{Privacy mechanisms for hiding the model, through adding noise or quantization, can be used in conjunction with the mechanism we propose here for protecting the gradient-Lipschitz constants. However, this is not the focus of this paper and we defer the analysis of such schemes to a future work.}.

 We propose Algorithm~\ref{Alg3} which is  a modification of Algorithm~\ref{Alg2} that makes  it privacy-preserving by adding an LDP mechanism on the gradient-Lipschitz constants. We define this mechanism below and refer to it as the  {\em {Gamma  mechanism}}.
 
  


\begin{defn}[Gamma mechanism]
\label{Gammadef}
The Gamma mechanism  takes as input the    gradient-Lipschitz constant $L_i$ of node $i$ and outputs  \begin{align*}
   \mathcal{R}(L_i)\sim{{Gamma  }}(\frac{L_{i}}{\theta},\,\theta),
\end{align*}
where $\theta >0$ is a noise parameter that controls the privacy level. 
\end{defn}

As a reminder, the probability density function of a   random variable $L$ distributed according to Gamma$(k,\theta)$ with a shape $k>0$ and scale $\theta>0$  is
 $$p_{{G,L }}\left(l\right)=\frac{1}{\varGamma\left(k\right)\theta^{k}}l^{k-1}e^{-\frac{l}{\theta}}, \text{  for } \,\, l>0, $$where
 $\Gamma\left(z\right)=\int_{0}^{\infty}u^{z-1}e^{-u}du  \text{\,\,\,\,for\,\,\,\,} z>0.$ 
 
 If $k$ is an integer, the Gamma distribution can be interpreted as the sum of $k$ exponential random variable with mean $\theta$ \cite{Gammawiki}.
 Computing the mean and the variance of  $\mathcal{R}(L_i)$, we get the mean $\mathbb{E}_{\mathcal{R}}\left[ \mathcal{R}(L_i)\right]=\frac{L_{i}}{\theta}\theta=L_{i},$ and the variance $v_{\mathcal{R}}\left[\mathcal{R}(L_{i})\right]=L_{i}\theta$  and thus $\theta$ determines  the noise level in the Gamma mechanism.


Accordingly, we obtain   Algorithm \ref{Alg3} that we call Private Weighted Random Walk SGD:


\begin{algorithm}[H]
   \caption{Private Weighted Random Walk SGD}
\begin{algorithmic}

  \STATE{ {\bfseries Initialization:} Initial node $i^{(1)}$ chosen uniformly at random from $[N]$, Initial model $w^{(1)}$ chosen uniformly at random from $\mathcal{W}$}.

   \STATE Every node $i$ computes $\mathcal{R}(L_i)$ and shares it along with $\deg(i)$ with its neighbors.

   \FOR{$k=1$ {\bfseries to} $T$}
    
   \STATE $ \hat{\nabla} f\left(w^{(k)}\right) = {\frac{\bar{L}}{L_{i^{(k)}}}}{\nabla}f_{i^{(k)}}\left(w^{(k)}\right)$
    \STATE $w^{(k+1)}={\bf{\Pi}}_{\mathcal{W}}\left(w^{(k)}-{\gamma^{(k)}}\hat{\nabla} f\left(w^{(k)}\right)\right)$
    
   \STATE Choose node $j$ uniformly at random from $\mathcal{N}\left(i^{(k)}\right).$ 
   \STATE Generate $p \sim U\left(0,\,1\right)$ where $U$ is the uniform distribution.
   \IF{$p\leq \min \, \left\{1,\,\frac {\mathcal{R}({L}_j)}{\mathcal{R}({L}_{i^{(k)}})}\frac{\deg\left(i^{(k)}\right)}{\deg\left(j\right)}\right\}$}
   \STATE $i^{(k+1)}\leftarrow j $
   \ELSE 
   \STATE $i^{(k+1)}\leftarrow i^{(k)}$
   \ENDIF
   \ENDFOR
   \STATE {\bfseries Return:} $w^{(T)}.$     
\end{algorithmic}
\label{Alg3}
\end{algorithm}

In this algorithm, each node $i$ applies the Gamma mechanism  to its constant $L_i$ only once and shares $\mathcal{R}(L_i)$ with its neighbors before the start of the random walk. These fixed values $\mathcal{R}(L_i)$'s are then used throughout the algorithm\footnote{Note that the alternative option of  generating and sharing a new value $\mathcal{R}(L_i)$ every time node $i$ is activated is vulnerable to an averaging attack that can be implemented by the neighbors of $i$.}.   The stationary distribution of the weighted random walk   of Algorithm~\ref{Alg3} with the noisy values $\mathcal{R}(L_i)$ is    $$\pi_{w, \mathcal{R}}(i) = \frac{\mathcal{R}~\left(L_{i}\right)}{\sum_{j}\mathcal{R}\left(L_{j}\right) },$$ 
and it follows the Beta distribution \cite{Gammawiki}, i.e., $\pi_{w, \mathcal{R}}(i) \sim\text{Beta}\left(\frac{L_{i}}{\theta},\,\frac{\sum_{j\neq  i}L_{j}}{\theta}\right).$ 
Note that the probability density function of the Beta distribution of a random variable $\Pi$, with parameters $\alpha,\,\beta >0$ is 
\begin{align*}
    p_{B,\Pi}(\pi)=\frac{\pi^{\alpha-1}\left(1-\pi\right)^{\beta-1}}{B\left(\alpha,\,\beta\right)}, \quad \text{for\,} \pi \in \left[0,\,1\right],
\end{align*}
where  
    \(B\left(\alpha,\,\beta\right)=\frac{\Gamma\left(\alpha\right)\Gamma\left(\beta\right)}{\Gamma\left(\alpha+\beta\right)}.\)
 
Therefore,  we get that the mean of $\pi_{w,\,\mathcal{R}}$ over the privatization randomness is the desired stationary distribution of importance sampling, i.e, 
\begin{align*}\mathbb{E}_{\mathcal{R}}\left[\ensuremath{\pi_{w,\,\mathcal{R}}}\right]=\pi_{w}(i),\,\,\,\,\forall i\in V.\end{align*}
We can see that the Beta distribution is well fitted to model the probability distribution of a probability \cite{aerinKim} where  the expected values of the stationary probabilities sums $1$, which justifies our choice for the Gamma noise over other traditional additive privacy schemes (e.g., Gaussian, Laplace) since it implies a Beta distribution on the ratio representing the stationary distribution.

Moreover, this implies that the  gradient estimate is unbiased at the stationary, that is 
\begin{align*}\mathbb{E}_{\mathcal{R}}\mathbb{E}_{\ensuremath{\pi_{w,\,\mathcal{R}}}}\left[\hat{\nabla}f_{i}\,|\,\mathcal{R}\left(L_{j}\right),\,j=1,\,...,\,N\right] & =\nabla f, 
\end{align*}
the proof of which can be found in Appendix~\ref{appendixPrivacyRate}. 

\begin{figure*}[t]

  \centering
  \subfigure[The achievable privacy parameter $\delta$ as a function of the privacy bound $\epsilon$.]{\includegraphics[scale=0.49]{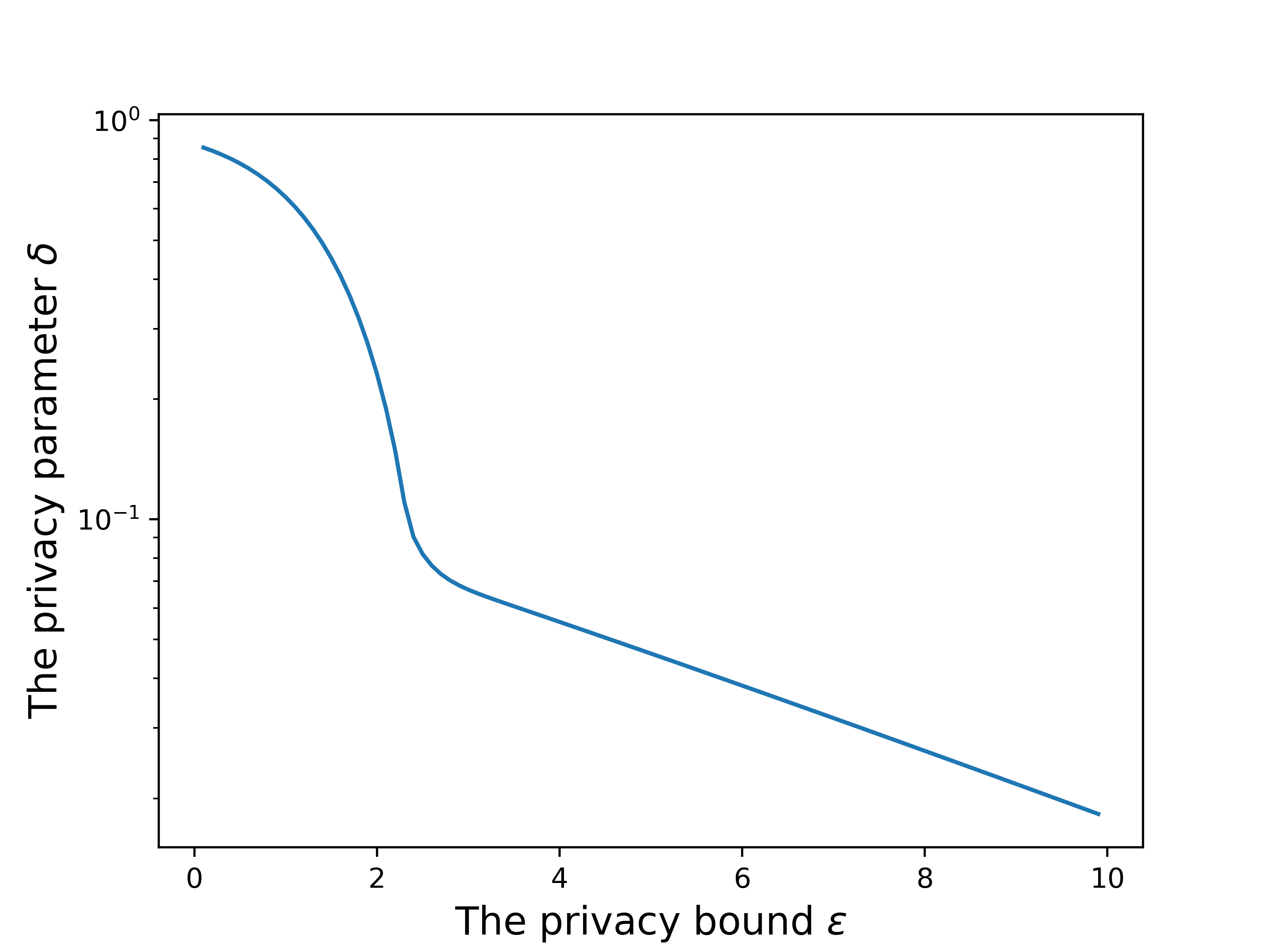}}
  \hspace{0.5em}
  \subfigure[The   Gamma noise parameter $\theta$ vs.\ the privacy bound $\epsilon$ for   $\delta = 0.06,\,0.08 \,\& \, 0.1$.]
  {\includegraphics[scale=0.49]{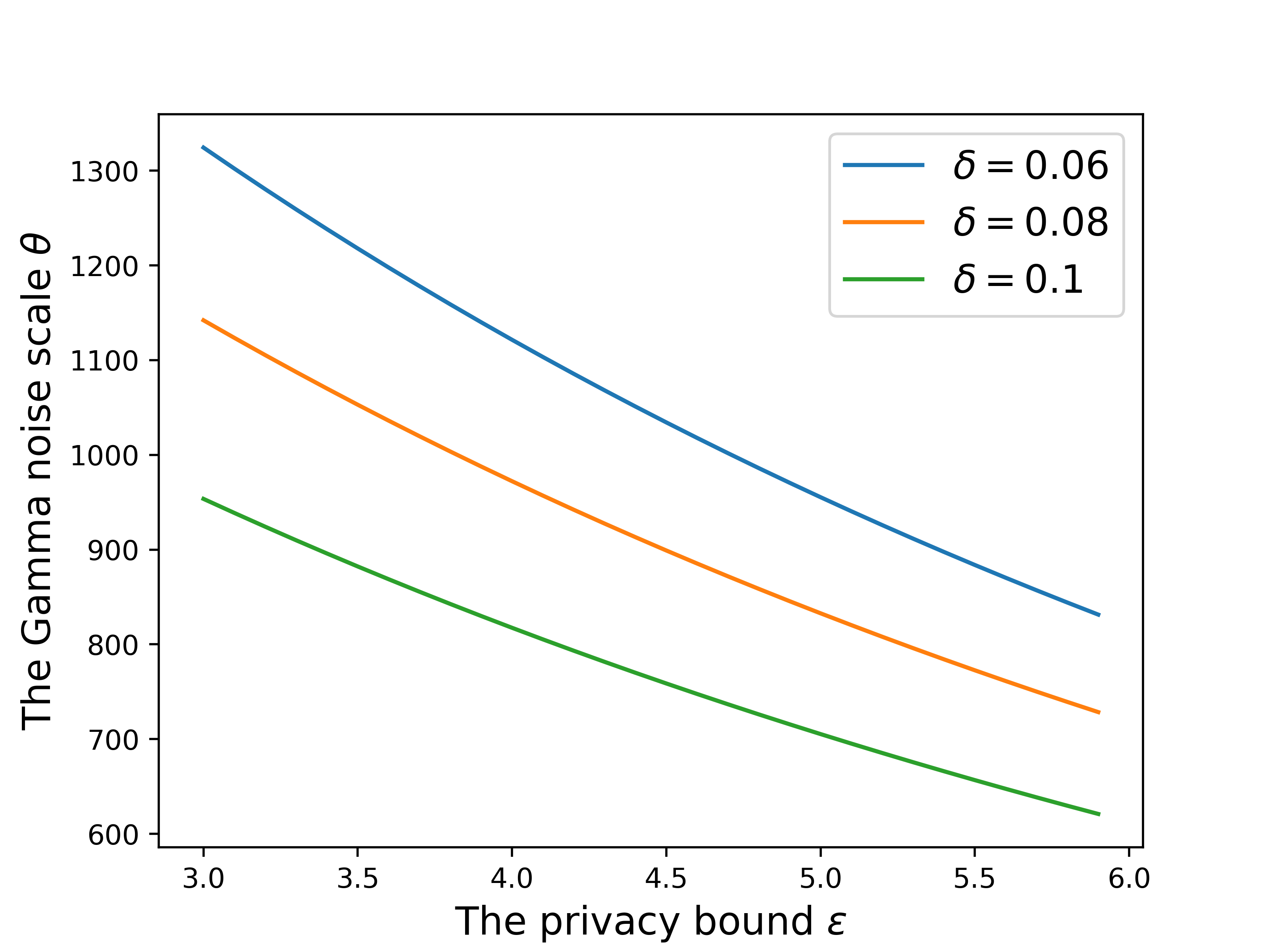}}
      \caption{The   tradeoffs  stated in Lemma~\ref{LemmaDP} between the privacy level,  quantified through $\epsilon$ and   $\delta$,  and  the noise level represented by $\theta$ for the Gamma mechanism.  \label{figurePrivacy}} 
\end{figure*}

The next Lemma states the tradeoff between the privacy level,  quantified through $\epsilon$ and   $\delta$,  and  the noise level represented by $\theta$ for the Gamma mechanism.
\begin{lemma}
Given $\epsilon\geq0$ , the Gamma mechanism is $\left(\epsilon,\,\delta\right)$ local differential private with parameter $\theta$ satisfying  

\begin{multline*} \delta\geq\max\left\{ 1-\frac{{IG}\left(\frac{\sup L}{\theta},\left(e^{\epsilon}\frac{\Gamma\left(\sup L/\theta\right)}{\Gamma\left(\inf L/\theta\right)}\right)^{\frac{\theta}{\sup L-\inf L}}\right)}{\Gamma\left(\sup L/\theta\right)} ,\frac{{IG}\left(\frac{\inf L}{\theta}\,,\,\left(\frac{1}{e^{\epsilon}}\frac{\Gamma\left(\sup L/\theta\right)}{\Gamma\left(\inf L/\theta\right)}\right)^{\frac{\theta}{\sup L-\inf L}} \right)}{\Gamma\left(\inf L/\theta  \right)} \right\}  \end{multline*}
where $IG\left(s,t\right):=\int_{0}^{t}u^{s-1}e^{-u}du$ is the incomplete gamma function.
\label{LemmaDP}
\end{lemma}
The proof of Lemma~\ref{LemmaDP} is in Appendix \ref{appendixPrivacyTradeoff}. In Fig.~\ref{figurePrivacy} shows plots of   the tradeoff  among the three parameters $\epsilon$, $\delta$ and $\theta$ descrbied in the above lemma.

\noindent \textbf{Convergence Analysis:} The proposed  Gamma mechanism applies the Gamma noise to the gradient-Lipschitz constants $L_i$'s only one time before the start of the algorithm. Therefore,
 finding the expression  of the rate of convergence in Algorithm~3, given the generated noisy Lipschitz constants  $\mathcal{R}\left(L_{i}\right)$, follows similar steps as the analysis of  non-private version of the algorithm and gives the following bound.
\begin{align*}&\mathbb{E}\left[f\left(\bar{w}^{(k)}\right)-f\left(w^{*}\right)|\,\mathcal{\mathcal{R}}\left(L_{i}\right),\,i=1,...,\,N\right] \\
& =\mathcal{O}\left(\frac{\max\left\{ \frac{\overline{\mathcal{\mathcal{R}}\left(L\right)}}{\inf\mathcal{\mathcal{R}}\left(L\right)}\sigma^{2},\,\left(\overline{\mathcal{\mathcal{R}}\left(L\right)}\right)^{2},\,\frac{1}{\ln1/\lambda_{P_{w,\,\mathcal{R}}}}\right\} }{k^{1-q}}\right). \label{private}\numberthis\end{align*}
To characterize the mean rate of convergence under  the Gamma mechanism we need to average the bound above over the  noise introduced by the mechanism.
This  involves finding the mean  of  the different terms in the right-hand-side of \eqref{private}, which is not straightforward given that the noise introduced by the Gamma mechanism is not additive. Instead, we analysis a modification of the mechanism that we call {\em Truncated Gamma} mechanism, which first implements the Gamma mechanism and then truncates the values  $\mathcal{R}\left(L_{i}\right)$ to a minimum and maximum  constants $L_{min}$ and $L_{max}$ ($L_{min}< L_{max}$).
By Proposition 2.1 in \cite{Dwork2014TheAF} on the  post-processing differential privacy immunity rule, we know that the Truncated Gamma mechanism achieves    the  privacy guarantees ($\epsilon, \delta$) of the non-truncated version and is more amenable to analysis as done below (as the expense of the tightness of the bound):
\begin{align*}\mathbb{E}_{\mathcal{R}}\left[\mathbb{E}\left[f\left(\bar{w}^{(k)}\right)-f\left(w^{*}\right)\right]\right]=O\left(\frac{\frac{L_{max}}{L_{min}}\sigma^{2}+\left(L_{\max}\right)^{2}}{k^{1-q}}\right) \label{latest} \numberthis .\end{align*}

We can see that the bound in \eqref{latest} benefits from the same asymptotic rate of convergence as the non-private version.

\section{Numerical Results}
\label{sectionV}
In this section, we present our numerical results applying Algorithms~\ref{Alg1}, \ref{Alg2} and \ref{Alg3} to a decentralized logistic regression problem and comparing it to   asynchronous Gossip SGD algorithm \cite{Ram2009AsynchronousGA}. We compare the Random Walk algorithms to the asynchronous Gossip to show the convergence speed-up with respect to the number of performed SGD iterations in the network. 
For fair comparison in terms of computation and communication, we assume an asynchronous Gossip algorithm where only one communication link gets activated at every iteration and the model on both side gets updated, exchanged and averaged.
Also, we run Algorithm~\ref{Alg3} using  the LDP privatization scheme based on the Gamma mechanism,  and we compare it to the additive Laplace mechanism widely used in the literature \cite{Dwork2014TheAF}.

 We take the graph $\mathcal{G}$ to be   an Erd\H os-R\'enyi graph on $N=100$ nodes with probability of connectivity $p=0.3$. 
  We assume that the labels $y_i\in\{-1,+1\}$.
  For the data with label $y_i=1$, we sample $x_i$ from $\mathcal{N}\left({{\mu}}, v \, \mathbb{I}_{d}\right)$ where $d$ is the feature dimension, $\mu$ is the mean, $v$ is the variance and  $ \mathbb{I}_{d}$ is the identity matrix of dimension $d$. For label $y_i=-1$, $x_i$ is sampled from $\mathcal{N}\left(-{\mu} ,v 
  \, \mathbb{I}_{d}\right)$. 

\noindent{\bf Loss function}: The averaged regularized cost function for logistic regression on the distributed data can be expressed as follow:
\begin{align}
    f\left(w\right)= \stackrel[i=1]{N}{\sum}\log\left(1+ \exp \left(-y_{i}x_{i}^{T}w\right)\right) + \frac{1}{2} \Vert w\Vert^2.
\end{align}

So, the local loss function at node $i$ is \begin{align}f_{i}\left(w\right)=N\log\left(1+\exp\left(-y_{i}x_{i}^{T}w\right)\right)+\frac{N}{2}\Vert w\Vert^{2}.\end{align}
Then, for such local loss function the gradient Lipschitz constant is $L_{i}=1+\frac{1}{4}N\left\Vert x_{i}\right\Vert _{2}^{2}$.

\begin{figure}[H]

  \centering
 
  \subfigure[High variance data regime:
  $v = 10$.]{\includegraphics[scale=0.5]{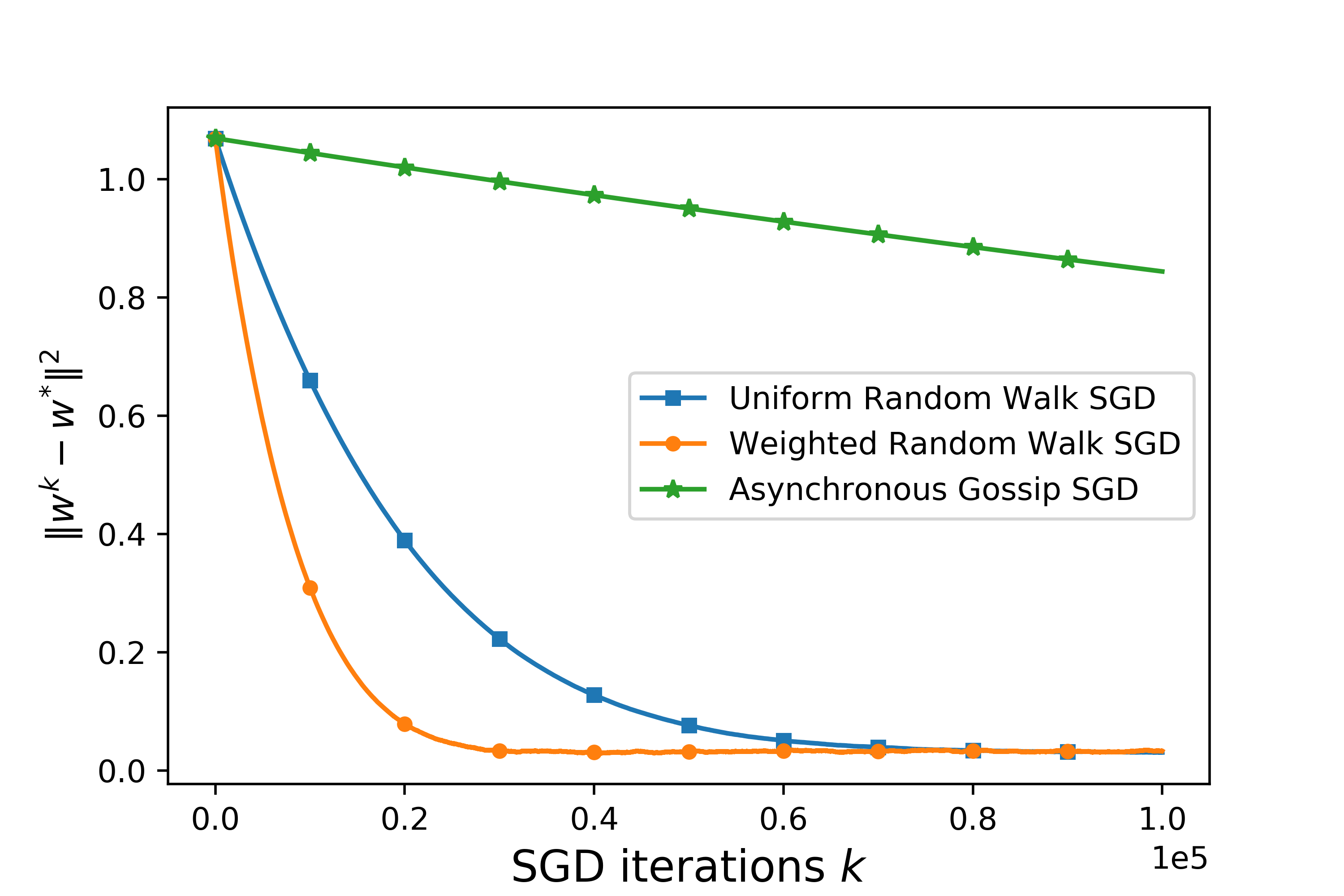}}
  \subfigure[Low variance data regime: $ v = 1$.]{\includegraphics[scale=0.5]{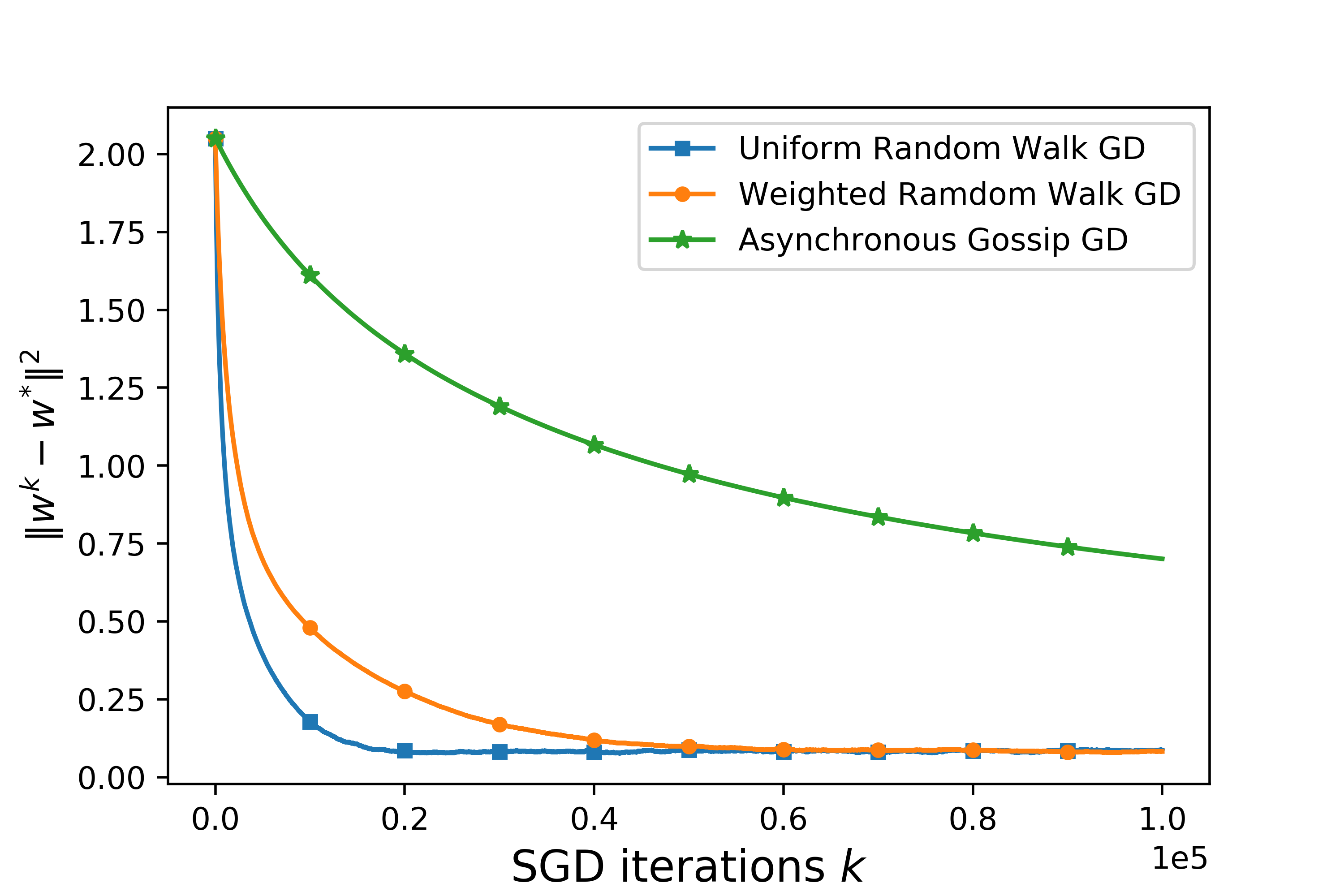}}

      \caption{Error on the model function of the SGD updates estimated by the Uniform Random Walk SGD (Algorithm~\ref{Alg1}), Weighted Random Walk SGD (Algorithm~\ref{Alg2}) and the asynchronous Gossip SGD on an Erd\H os-R\'enyi graph of $100$ nodes and probability of connectivity $p=0.3$.}
       \label{FigSGD}
\end{figure}

\noindent{\bf Random Walk algorithms vs the Gossip algorithm:} We consider two data regimes for our simulations: High variance and Low variance regimes.
For both regimes, the random walk based algorithms outperform the Gossip algorithm as shown in Figure~\ref{FigSGD}. For the high variance data regime where $v=10$, the term $\bar{L}$ in the convergence rate dominates, and the weighted sampling gives better convergence. While in the low variance data regime where $v =1$, the Uniform Random Walk SGD is outperforming the Weighted Random Walk SGD.

\noindent{\bf Private Random Walk algorithms:} For high variance regime, we simulate the convergence of Algorithm~\ref{Alg3} that used  the  Gamma mechanism. We compare its performance to a variant that uses     an additive  Laplacian noise,   \cite{Dwork2014TheAF} for $\epsilon = 3$ and $\delta \leq 0.03$ with same noise variance as the Gamma. Our results show that the Gamma mechanism outperforms the Laplacian mechanism .
\begin{figure}[H]
  \centering
  \subfigure{\includegraphics[scale=0.5]{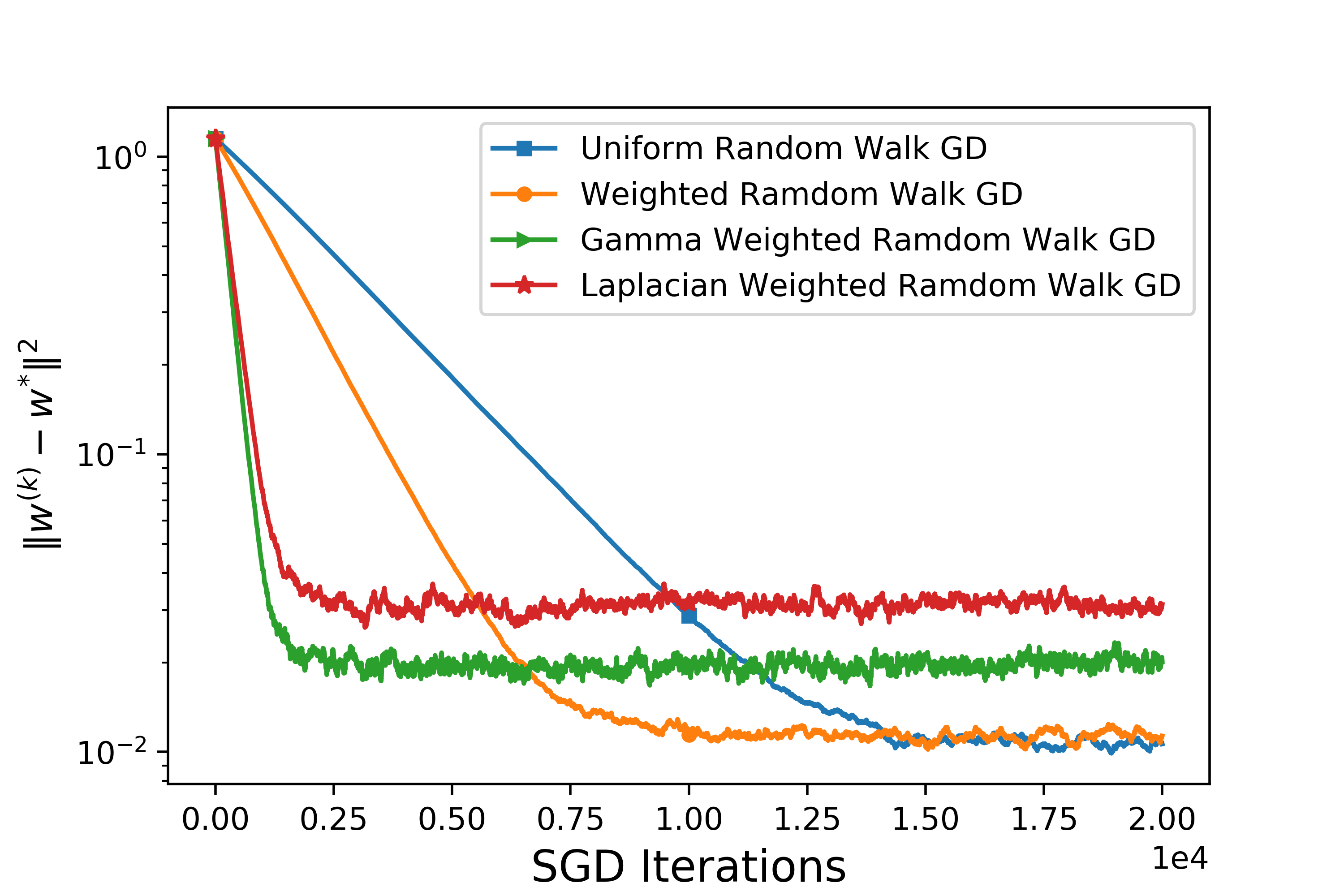}}

      \caption{Error of the model estimated by the Uniform Random Walk SGD, $(3,\,0.03)$-LDP Gamma Weighted Random walk (Algorithm~\ref{Alg3})  and $(3,\,0)$-LDP Laplace Weighted Random walk on an Erd\H os-R\'enyi graph of $20$ nodes and probability of connectivity $p=0.3$.}
    
\end{figure}

\section{Conclusion}
\label{sectionVI}
In this paper, we study a decentralized learning problem where the data is distributed over the nodes of a graph.    We propose   decentralized learning algorithms based on     random walk.   
To    speed up the convergence, we propose a weighted random walk algorithm where the nodes get sampled depending on the importance of their data measured through the gradient-Lipschitz constant. A weighted random walk requires every node to share its gradient-Lipschitz constant with the neighbor nodes, which creates a privacy vulnarability. We propose a local differential privacy mechanism that we call  Gamma mechanism to address the privacy concern, and   give the tradeoff between the privacy parameters and the Gamma noise parameter. We also presented numerical results on the convergence of the proposed algorithms. 
 As for future work, we think it is interesting  
  to investigate different importance sampling measures for the weighted random walk. 
 
 
\bibliography{biblo.bib}
\bibliographystyle{IEEEtran}

\newpage
\,
\newpage 
\appendices

\section{Proof of Theorem~\ref{tUniform}:  Convergence Rate of Algorithm~\ref{Alg1}}

\label{Appendix1}

First, we present preliminary results that will be used in the main proof later.

\begin{lemma} [Lipschitzness]
If $f_i$ is a convex function on an open subset $\Omega \subseteq \mathbb{R}$, then for a closed bounded subset $\mathcal{W} \subset \Omega$, there exists a constant $D_i\geq0$, such that, for any $w_1,\,w_2 \in \mathcal{W}$,
\begin{align*}
\left|f_{i}\left(w_{1}\right)-f_{i}\left(w_{2}\right)\right|\leq D_i\left\Vert w_1-w_2\right\Vert_2.   \end{align*}
\label{Lemma 3}
We define $D=\underset{i\in V}{\sup } D_i$. Therefore,
$$\left|f_{i}\left(w_{1}\right)-f_{i}\left(w_{2}\right)\right|\leq D\left\Vert w_1-w_2\right\Vert_2. $$
\end{lemma}
\noindent A proof for Lemma~\ref{Lemma 3} can be found in \cite{ar}.
\begin{cor}[Boundedness of the gradient]
If $f_i$ is a convex function on  $ \mathbb{R}$, then for a closed bounded subset $\mathcal{W} \subset \mathbb{R}$,
\(\left\Vert \nabla f_i(w) \right\Vert_2\leq D, \,\,\,\,\, \forall w\in \mathcal{W}.\)
\begin{proof}
Taking $v = w+\nabla f_i (w)$, 
\begin{align*}D\left\Vert \nabla f_{i}(w)\right\Vert _{2} =D\left\Vert v-w\right\Vert _{2}
  \overset{(a)}{\geq}|f_{i}(v)-f_{i}(w)| 
  \overset{(b)}{\geq}\left\langle \nabla f_{i}\left(w\right),\nabla f_{i}\left(w\right)\right\rangle 
 =\left\Vert \nabla f_{i}(w)\right\Vert _{2}^{2}.
\end{align*}
(a) follows Lemma~\ref{Lemma 3} and (b) follows from $f_i$ being convex.

\end{proof}

\label{cor1}
\end{cor}

Next we give some results we need on the  Markov chain. We remind the reader that $\pi$ is the stationary distribution, $P$ is the transition matrix and $P^k$ is the $k^{th}$ power of matrix P. We refer to   $i^{th}$ row of a matrix $P$ by $P(i,:)$.

\begin{lemma} [Convergence of Markov Chain \cite{Levin}] 
Under Assumption~\ref{assumption2},  we have
\begin{align*}
    \underset{i}{\max}\left\Vert \pi-P^{k}\left(i,:\right)\right\Vert \leq C\lambda_{P}^{k} 
\end{align*}
 for $k > K_P$, where
$K_P$ is a constant that depends on $\lambda_{P}$ and $\lambda_2(P)$ and $C$ is a constant that depends on the Jordan canonical form of $P$.

\end{lemma}


Next, we start the proof for Theorem~\ref{tUniform} following the same reasoning in \cite{MCGD} but with the inclusion of the gradient-Lipschitz assumption as in Assumption \ref{assumption1}.
\begin{align*}  \left\Vert w^{(k+1)}-w^{*}\right\Vert _{2}^{2}  & =\left\Vert {\bf {\bf {\Pi}}_{\mathcal{W}}}\left(w^{(k)}-\gamma^{(k)}\nabla f_{i^{(k)}}\left(w^{(k)}\right)\right)-{\bf {\bf {\Pi}}_{\mathcal{W}}}\left(w^{*}\right)\right\Vert _{2}^{2}\\
 & \overset{\left(a\right)}{\leq}\left\Vert w^{(k)}-\gamma^{(k)}\nabla f_{i^{(k)}}\left(w^{(k)}\right)-w^{*}\right\Vert _{2}^{2}\\
 & =\left\Vert w^{(k)}-w^{*}\right\Vert _{2}^{2}-2\gamma^{(k)}\left\langle w^{(k)}-w^{*},\nabla f_{i^{(k)}}\left(w^{(k)}\right)\right\rangle \\
 & \,\,\,\,\,+\left(\gamma^{(k)}\right){}^{2}\left\Vert \nabla f_{i^{(k)}}\left(w^{(k)}\right)\right\Vert _{2}^{2}\\
 & =\left\Vert w^{(k)}-w^{*}\right\Vert _{2}^{2}-2\gamma^{(k)}\left\langle w^{(k)}-w^{*},\nabla f_{i^{(k)}}\left(w^{(k)}\right)\right\rangle \\
 & \,\,\,\,\,+\left(\gamma^{(k)}\right){}^{2}\left\Vert \nabla f_{i^{(k)}}\left(w^{(k)}\right)-\nabla f_{i^{(k)}}\left(w^{*}\right)+\nabla f_{i^{(k)}}\left(w^{*}\right)\right\Vert _{2}^{2}\\
 & \overset{\left(b\right)}{\leq}\left\Vert w^{(k)}-w^{*}\right\Vert _{2}^{2}-2\gamma^{(k)}\left\langle w^{(k)}-w^{*},\nabla f_{i^{(k)}}\left(w^{(k)}\right)\right\rangle \\
 & \,\,\,\,\,+2{(\gamma^{(k)})}^{2}\left\Vert \nabla f_{i^{(k)}}\left(w^{(k)}\right)-\nabla f_{i^{(k)}}\left(w^{*}\right)\right\Vert _{2}^{2}\\
 & \,\,\,\,\,+2{(\gamma^{(k)})}^{2}\left\Vert \nabla f_{i^{(k)}}\left(w^{*}\right)\right\Vert _{2}^{2}.\numberthis
\end{align*}

\noindent $\left(a\right)$ follows from $\mathcal{W}$ being a convex closed set, so one can apply nonexpansivity theorem \cite [Fact E.9.0.0.5]{convex},
$\left(b\right)$ follows from Jensen's inequality applied to the squared norm.
\\
Using Lemma~\ref{Lemma 3} and the convexity of the functions $f_i$, we get
\begin{align*}  \left\Vert w^{(k+1)}-w^{*}\right\Vert _{2}^{2}  & \overset{\left(a\right)}{\leq}\left\Vert w^{(k)}-w^{*}\right\Vert _{2}^{2}-2\gamma^{(k)}\left\langle w^{(k)}-w^{*},\nabla f_{i^{(k)}}\left(w^{(k)}\right)\right\rangle \\
 & +2\left(\gamma^{(k)}\right){}^{2}L_{i^{(k)}}^{2}\left\Vert w^{(k)}-w^{*}\right\Vert _{2}^{2}+2\left(\gamma^{(k)}\right){}^{2}\left\Vert \nabla f_{i^{(k)}}\left(w^{*}\right)\right\Vert _{2}^{2}\\
 & \overset{\left(b\right)}{\leq}\left\Vert w^{(k)}-w^{*}\right\Vert _{2}^{2}-2\gamma^{(k)}\left\langle w^{(k)}-w^{*},\nabla f_{i^{(k)}}\left(w^{(k)}\right)\right\rangle \\
 &    +2{(\gamma^{(k)})}^{2}\left(\sup L\right)^{2}\left\Vert w^{(k)}-w^{*}\right\Vert _{2}^{2}%
+  2{(\gamma^{(k)})}^{2}\left\Vert \nabla f_{i^{}}\left(w^{*}\right)\right\Vert _{2}^{2}.\numberthis\label{eq3}
\end{align*}

\noindent  $\left(a\right)$ follows from the Lipschitzness Lemma, 
$\left(b\right)$ follows by bounding by the $\sup L$.

\noindent For the next we use the convexity of $f_i$,
\begin{align*}   \left\Vert  w^{(k+1)}-w^{*}\right\Vert^2_2  &\leq\left\Vert w^{(k)}-w^{*}\right\Vert _{2}^{2}-2\gamma^{(k)}\left(f_{i^{(k)}}\left(w^{(k)}\right)-f_{i^{(k)}}\left(w^{*}\right)\right)
 \\
 &  
+2{(\gamma^{(k)})}^{2}\left(\sup L\right)^{2}\left\Vert w^{(k)}-w^{*}\right\Vert _{2}^{2}
 \\
 & 
+2{(\gamma^{(k)})}^{2}\left\Vert \nabla f_{i^{(k)}}\left(w^{*}\right)\right\Vert _{2}^{2}.
\label{eq:4}
\numberthis
\end{align*}

\noindent By re-arranging \eqref{eq:4}, we come to

\begin{align*}   \gamma^{(k)}\left(f_{i^{(k)}}\left(w^{(k)}\right)-f_{i^{(k)}}\left(w^{*}\right)\right) 
 & \leq\frac{1}{2}\left(\left\Vert w^{(k)}-w^{*}\right\Vert _{2}^{2}-\left\Vert w^{(k+1)}-w^{*}\right\Vert _{2}^{2}\right)
\\ & \quad +{(\gamma^{(k)})}^{2}\left(\sup L\right)^{2}\left\Vert w^{(k)}-w^{*}\right\Vert _{2}^{2} \\ & \quad
 +{(\gamma^{(k)})}^{2}\left\Vert \nabla f_{i^{(k)}}\left(w^{*}\right)\right\Vert _{2}^{2}. \label{eq:8}
 \numberthis
\end{align*}

\noindent Now summing \eqref{eq:8} over $k$ and using Assumption~\ref{assumption1} and the boundness of $\mathcal{W}$,
\begin{align*}\sum_{k}\gamma^{(k)}\left(f_{i^{(k)}}\left(w^{(k)}\right)-f_{i^{(k)}}\left(w^{*}\right)\right) & \leq\frac{1}{2}\left\Vert w^{(0)}-w^{*}\right\Vert _{2}^{2}+\left(\sup L\right)^{2}\sum_{k}(\gamma^{(k)})^{2}\left\Vert w^{(k)}-w^{*}\right\Vert _{2}^{2}\\
 & +\sum_{k}(\gamma^{(k)})^{2}\left\Vert \nabla f_{i^{(k)}}\left(w^{*}\right)\right\Vert _{2}^{2}<\infty.\label{eqn}\numberthis
\end{align*}

We can see that the previous result shows the dependency on $\sup L$.
\begin{align*}  \gamma^{(k)}\mathbb{E}\left[f_{j^{(k)}}\left(w^{\left(k-T^{\left(k\right)}\right)}\right)-f_{j^{(k)}}\left(w^{(k)}\right)\right]   &  \overset{(a)}{\leq}D\gamma^{(k)}\mathbb{E}\left\Vert w^{\left(k-T^{\left(k\right)}\right)}-w^{(k)}\right\Vert \\
 &  \overset{(b)}{\leq}D\gamma^{(k)}\mathbb{E}\left(\sum_{n=k-T^{\left(k\right)}}^{k-1}\left\Vert w^{\left(n+1\right)}-w^{\left(n\right)}\right\Vert \right)\\
 &  \overset{(c)}{\leq}D\gamma^{(k)}\sum_{n=k-T^{\left(k\right)}}^{k-1}\mathbb{E}\left(\left\Vert w^{\left(n+1\right)}-w^{\left(n\right)}\right\Vert \right)\\
 &   \overset{(d)}{\leq}D^{2}\gamma^{(k)}\sum_{n=k-T^{\left(k\right)}}^{k-1}\gamma^{\left(n\right)}\\
 &  \overset{(e)}{\leq}\frac{D^{2}}{2}\sum_{n=k-T^{\left(k\right)}}^{k-1}\left({(\gamma^{\left(n\right)})}^{2}+{(\gamma^{\left(k\right)})}^{2}\right)\\
 &  \overset{}{\leq}\frac{D^{2}}{2}T^{\left(k\right)}{(\gamma^{\left(k\right)})}^{2}+\frac{D^{2}}{2}\sum_{n=k-T_{k}}^{k-1}{(\gamma^{\left(n\right)})}^{2}.
\end{align*}

(a) follows from Lemma~\ref{Lemma 3}, (b) using triangle inequality, (c) using linearity of expectation, (d) follows Corollary~\ref{cor1} and (e) follows from the Cauchy–Schwarz inequality.

\noindent Now taking the summation over $k$:
 \begin{align*}  \sum_{k}\gamma^{(k)}\mathbb{E}  \left[f_{j^{(k)}}\left(w^{(k-T^{(k)})}\right)-f_{j^{(k)}}\left(w^{(k)}\right)\right]
  \leq\sum_{k}\frac{D^{2}}{2}T_k{(\gamma^{(k)})}^{2}+\frac{D^{2}}{2}\sum_{k}\sum_{n=k-T_k}^{k-1}{(\gamma^{(n)})}^{2}.
 \label{bound1}\numberthis
 \end{align*}
By simply using the assumption on the step size summability, the result is as follows:
      \begin{align*}
      \label{FirstEq}
      \sum_{k=K}^{\infty}\sum_{n=k-T^{(k)}}^{k-1}\left(\gamma^{\left(n\right)}\right)^{2}
      \leq\sum_{k=K}^{\infty}T^{\left(k\right)}\left(\gamma^{(k)}\right)^{2}\leq\frac{2}{\ln\left(1/\lambda_{P}\right)}\sum_{k=K}^{\infty}\ln k.\left(\gamma^{(k)}\right)^{2}<\infty. \numberthis \end{align*}

\begin{align*} & \mathbb{E}_{j^{(k)}}\left[f_{j^{(k)}}\left(w^{(k-T^{\left(k\right)})}\right)-f_{j^{(k)}}\left(w^{*}\right)|\,X_{0},\,X_{1},\,...,\,X_{k-T^{\left(k\right)}}\right]\\
 & \overset{}{=}\sum_{i=1}^{N}\left(f_{i}\left(w^{(k-T^{\left(k\right)})}\right)-f_{i}\left(w^{*}\right)\right)P\left(j^{(k)}=i\,|\,X_{0},\,X_{1},\,...,\,X_{k-T^{\left(k\right)}}\right)\\
 & \overset{(a)}{=}\sum_{i=1}^{N}\left(f_{i}\left(w^{(k-T^{\left(k\right)})}\right)-f_{i}\left(w^{*}\right)\right)P\left(j^{(k)}=i\,|\,X_{k-T^{\left(k\right)}}\right)\\
 & \overset{}{=}\sum_{i=1}^{N}\left(f_{i}\left(w^{(k-T^{\left(k\right)})}\right)-f_{i}\left(w^{*}\right)\right)P^{t_{k}}\left[X_{k-T^{\left(k\right)}}\,|\,j^{(k)}=i\right]\\
 & \overset{(b)}{\geq}\left(f\left(w^{(k-T^{\left(k\right)})}\right)-f\left(w^{*}\right)\right)-\frac{N}{2k}.
\end{align*}

(a) using Markov property and (b) using Lemma 1 in \cite{MCGD}. \\

Next, we get a bound on $$  \sum_{k}\gamma^{\left(k\right)}\mathbb{E}\left[f\left(w^{(k)}\right)-f\left(w^{(k-T^{\left(k\right)})}\right)\right].$$
\begin{align*}\gamma^{(k)}\mathbb{E}\left[f\left(w^{\left(k-T_{k}\right)}\right)-f\left(w^{(k)}\right)\right] & \overset{(a)}{\leq}ND\gamma^{(k)}\mathbb{E}\left\Vert w^{\left(k-T_{k}\right)}-w^{(k)}\right\Vert \\ &  \overset{(b)}{\leq}ND\gamma^{(k)}\mathbb{E}\left(\sum_{n=k-T_{k}}^{k-1}\left\Vert w^{\left(n+1\right)}-w^{\left(n\right)}\right\Vert \right)\\ & \overset{(c)}{\leq}ND\gamma^{(k)}\sum_{n=k-T_{k}}^{k-1}\mathbb{E}\left(\left\Vert w^{\left(n+1\right)}-w^{\left(n\right)}\right\Vert \right) \\ & \overset{(d)}{\leq}ND^{2}\gamma^{(k)}\sum_{n=k-T_{k}}^{k-1}\gamma^{\left(n\right)} \\ &  \overset{(e)}{\leq}\frac{ND^{2}}{2}\sum_{n=k-T_{k}}^{k-1}\left({(\gamma^{\left(n\right)})}^{2}+{(\gamma^{\left(k\right)})}^{2}\right) \\ &  \overset{}{\leq}\frac{ND^{2}}{2}T_{k}{(\gamma^{\left(k\right)})}^{2}+\frac{ND^{2}}{2}\sum_{n=k-T_{k}}^{k-1}{(\gamma^{\left(n\right)})}^{2}.
\end{align*}
(a) follows from Lemma 4, (b) using triangle inequality, (c) using linearity of expectation, (d) follows Corollary 1 and (e) follows from the Cauchy–Schwarz inequality.
The upper bound summability over $k$ follows from previous discussion in equation \eqref{FirstEq}.

\noindent Combining with the results in \eqref{eqn} and \eqref{FirstEq} , we get 
\begin{align*}  \sum_{k}\gamma^{(k)}\mathbb{E}\left[f\left(w^{(k-T_k)}\right)-f\left(w^{*}\right)\right]   \leq C_{1}\sigma^{2} +C_{2}\left(\sup L\right)^{2}+C_{3}\left(\frac{1}{\ln\left(1/\lambda_{P}\right)}\right).
\end{align*}

Finally,
\begin{align*} &  \mathbb{E}\left(f\left(\bar{w}^{(k)}\right)-f\left(w^{*}\right)\right)   =O\left(\frac{\max\left(\sigma^{2},\,\left(\sup L\right)^{2},\,\frac{1}{\ln\left(1/\lambda_{P}\right)}\right)}{\left(\stackrel[j=1]{k}{\sum}\gamma^{(k)}\right)}\right) =O\left(\frac{\max\left(\sigma^{2},\,\left(\sup L\right)^{2},\,\frac{1}{\ln\left(1/\lambda_{P}\right)}\right)}{k^{1-q}}\right).\numberthis\label{eq9}
\end{align*}

\section{ Proof of Theorem \ref{tWeighted} } 

\label{AppendixProofWeighted}

To prove Theorem \ref{tWeighted}, we scale the local losses  $f_i\left(w\right)$ in order to keep the same global loss under the new weighted sampling average is
\begin{align}
f_{i,\,w}\left(w\right)=\frac{f_{i}\left(w\right)}{\frac{L_{i}}{\bar{L}}}=\frac{\bar{{L}}}{L_{i}}f_{i}\left(w\right).
\end{align}

For the next, we compute the gradient Lipschitz constant for the weighted loss function which is $ L_{i,\,w}=\frac{L_{i}}{\frac{L_{i}}{\bar{L}}}=\bar{{L}}.$

Here, we get 
$\sup L_w:=\underset{i}{\max}L_{i,\,w}=\bar{{L}}.$
After we compute the new residual quantity that also contributes to the rate of convergence:
\begin{align*}
    \sum_{i=1}^{N}\frac{L_{i}}{N\bar{{L}}}\left\Vert \nabla f_{i,\,w}\left(w^{*}\right)\right\Vert _{2}^{2} 	  & =\sum_{i=1}^{N}\frac{L_{i}}{N\bar{{L}}}\left(\frac{\bar{L}}{L_{i}}\right)^{2}\left\Vert \nabla f_{i}\left(w^{*}\right)\right\Vert _{2}^{2}
 \\ &	=\sum_{i=1}^{N}\frac{1}{N}\left(\frac{\bar{L}}{L_{i}}\right)\left\Vert \nabla f_{i}\left(w^{*}\right)\right\Vert _{2}^{2}
\\ & 	\leq\frac{\bar{L}}{\inf L}\left(\sum_{i=1}^{N}\frac{1}{N}\left\Vert \nabla f_{i}\left(w^{*}\right)\right\Vert _{2}^{2}\right).
\end{align*}

\section{Private Random Walk SGD   }
\label{appendixPrivacyRate}

Using the Gamma noise on the gradient-Lipschitz constants in the weighted random walk ends up giving unbiased gradient estimate at the stationary regime:
\begin{align*}\mathbb{E}_{\mathcal{R}}\mathbb{E}_{\ensuremath{\pi_{w,\,\mathcal{R}}}}\left[\hat{\nabla}f_{i}\,|\,\mathcal{R}\left(L_{j}\right),\,j=1,\,...,\,N\right] & =\nabla f.\end{align*}
\begin{proof}
\begin{align*}  \mathbb{E}_{\mathcal{R}}\mathbb{E}_{\ensuremath{\pi_{w,\,\mathcal{R}}}}\left[\hat{\nabla}f_{i}\,|\,\mathcal{R}\left(L_{j}\right),\,j=1,\,...,\,N\right]  &  =\mathbb{E}_{\mathcal{R}}\mathbb{E}_{\ensuremath{\pi_{w,\,\mathcal{R}}}}\left[\frac{\bar{L}}{L_{i}}\nabla f_{i}|\,\mathcal{R}\left(L_{j}\right),\,j=1,\,...,\,N\right]\\
 &  = \sum_{i}\mathbb{E}_{\mathcal{R}}\left[\frac{\mathcal{R}\left(L_{i}\right)}{\sum_{j}\mathcal{R}\left(L_{j}\right)}\frac{\sum_{j}L_{j}}{NL_{i}}\nabla f_{i}\right]\\
 &   =\mathbb{E}_{i}\left[\nabla f_{i}\right]=\nabla f.
\end{align*}
\end{proof}



\section{Proof of Lemma~\ref{LemmaDP}}
\label{appendixPrivacyTradeoff}
We will use the following property for $(\epsilon,\, \delta)$  differentially mechanism. 
\begin{lemma}\cite{balle}

A mechanism is $(\epsilon, \delta)$-differentially private if,  for any $L,\,L'$, and $Z$ a random variable with the same distribution as $\mathcal{R}(L)$, we have \begin{align*}P(\log \frac{P_{\mathcal{R}\left(L\right)}\left(z\right)}{P_{\mathcal{R}\left(L'\right)}\left(z\right)}\geq\epsilon)\leq\delta.\end{align*}
\end{lemma}
In our case, $Z \sim Gamma (L,\,\theta)$.

Now, for the proof of Lemma \ref{LemmaDP}, taking $L,\, L'$ and a received output $z \in \mathbb{R}$, we have

\begin{align*}\frac{P_{\mathcal{R}\left(L\right)}\left(z\right)}{P_{\mathcal{R}\left(L'\right)}\left(z\right)}
 =\frac{\frac{1}{\Gamma\left(L/\theta\right)\theta^{\frac{L}{\theta}}}z^{\frac{L}{\theta}-1}e^{\frac{z}{\theta}}}{\frac{1}{\Gamma\left(L'/\theta\right)\theta^{\frac{L'}{\theta}}}z^{\frac{L'}{\theta}-1}e^{\frac{z}{\theta}}} =\frac{\Gamma\left(L'/\theta\right)}{\Gamma\left(L/\theta\right)}\theta^{\frac{L'-L}{\theta}}z^{\frac{L-L'}{\theta}}.
\end{align*}


Then, taking a random variable $Z$ on $z$, where $Z \sim Gamma(L,\, \theta)$: 
\begin{align*}& \left(\frac{Z}{\theta}\right)^{\frac{L-L'}{\theta}}\sim {Generalized \, Gamma}\left(p=\frac{\theta}{L-L'},\,d=\frac{L}{L-L'},\,1\right), \\ & \quad \text{when}\, L>L'.\end{align*}
As a reminder, the probability density function of the Generalized Gamma function on a random variable $Y$ is 
\begin{align*}
  p_{GG,\,Y}(y)=  \frac{p/a^{d}}{\Gamma\left(d/p\right)}y^{d-1}e^{-\left(y/a\right)^{p}}\qquad y>0,\quad a,d,p>0.
\end{align*}

When  $L>L'$, we have 
\begin{align*}P\left(\frac{\Gamma\left(L'/\theta\right)}{\Gamma\left(L/\theta\right)}\theta^{\frac{L'-L}{\theta}}Z^{\frac{L-L'}{\theta}}\geq e^{\epsilon}\right) & =P\left(\left(\frac{Z}{\theta}\right)^{\frac{L-L'}{\theta}}\geq e^{\epsilon}\frac{\Gamma\left(L/\theta\right)}{\Gamma\left(L'/\theta\right)}\right)\\
 & =1-P\left(\left(\frac{Z}{\theta}\right)^{\frac{L-L'}{\theta}}\leq e^{\epsilon}\frac{\Gamma\left(L/\theta\right)}{\Gamma\left(L'/\theta\right)}\right)\\
 & =1-\frac{IG\left(\frac{L}{\theta},\left(e^{\epsilon}\frac{\Gamma\left(L/\theta\right)}{\Gamma\left(L'/\theta\right)}\right)^{\frac{\theta}{L'-L}}\right)}{\Gamma\left(L/\theta\right)}\\
 & \leq\delta,
\end{align*}

where $IG$ is the lower incomplete gamma function.

When  $L<L'$, we have 
\begin{align*}P\left(\frac{\Gamma\left(L'/\theta\right)}{\Gamma\left(L/\theta\right)}\theta^{\frac{L'-L}{\theta}}Z^{\frac{L-L'}{\theta}}\geq e^{\epsilon}\right) & =P\left(\left(\frac{Z}{\theta}\right)^{\frac{L'-L}{\theta}}\leq\frac{\Gamma\left(L'/\theta\right)}{\Gamma\left(L/\theta\right)e^{\epsilon}}\right)\\
 & =\frac{IG\left(\frac{L}{\theta},\left(\frac{\Gamma\left(L'/\theta\right)}{\Gamma\left(L/\theta\right)e^{\epsilon}}\right)^{\frac{\theta}{L'-L}}\right)}{\Gamma\left(L/\theta\right)}\\
 & \leq\delta.
\end{align*}


\bibliographystyle{IEEEtran}
\bibliography{biblo.bib}

\begin{thebibliography}{10}
\providecommand{\url}[1]{#1}
\csname url@samestyle\endcsname
\providecommand{\newblock}{\relax}
\providecommand{\bibinfo}[2]{#2}
\providecommand{\BIBentrySTDinterwordspacing}{\spaceskip=0pt\relax}
\providecommand{\BIBentryALTinterwordstretchfactor}{4}
\providecommand{\BIBentryALTinterwordspacing}{\spaceskip=\fontdimen2\font plus
\BIBentryALTinterwordstretchfactor\fontdimen3\font minus
  \fontdimen4\font\relax}
\providecommand{\BIBforeignlanguage}[2]{{%
\expandafter\ifx\csname l@#1\endcsname\relax
\typeout{** WARNING: IEEEtran.bst: No hyphenation pattern has been}%
\typeout{** loaded for the language `#1'. Using the pattern for}%
\typeout{** the default language instead.}%
\else
\language=\csname l@#1\endcsname
\fi
#2}}
\providecommand{\BIBdecl}{\relax}
\BIBdecl

\bibitem{Ram2009AsynchronousGA}
S.~S. Ram, A.~Nedic, and V.~V. Veeravalli, ``Asynchronous gossip algorithms for
  stochastic optimization,'' \emph{2009 International Conference on Game Theory
  for Networks}, pp. 80--81, 2009.

\bibitem{MCGD}
T.~Sun, Y.~Sun, and W.~Yin, ``On markov chain gradient descent,'' in
  \emph{NeurIPS}, 2018.

\bibitem{Needell}
D.~Needell, R.~Ward, and N.~Srebro, ``{Stochastic Gradient Descent, Weighted
  Sampling, and the Randomized Kaczmarz algorithm},'' in \emph{Advances in
  Neural Information Processing Systems 27}, Z.~Ghahramani, M.~Welling,
  C.~Cortes, N.~D. Lawrence, and K.~Q. Weinberger, Eds.\hskip 1em plus 0.5em
  minus 0.4em\relax Curran Associates, Inc., 2014, pp. 1017--1025.

\bibitem{Johansson2007ASP}
B.~Johansson, M.~Rabi, and M.~Johansson, ``A simple peer-to-peer algorithm for
  distributed optimization in sensor networks,'' \emph{2007 46th IEEE
  Conference on Decision and Control}, pp. 4705--4710, 2007.

\bibitem{Johansson2009ARI}
------, ``A randomized incremental subgradient method for distributed
  optimization in networked systems,'' \emph{SIAM J. Optim.}, vol.~20, pp.
  1157--1170, 2009.

\bibitem{Nedic2009DistributedSM}
A.~Nedic and A.~Ozdaglar, ``Distributed subgradient methods for multi-agent
  optimization,'' \emph{IEEE Transactions on Automatic Control}, vol.~54, pp.
  48--61, 2009.

\bibitem{Wai2018SUCAGSU}
H.~Wai, N.~Freris, A.~Nedic, and A.~Scaglione, ``Sucag: Stochastic unbiased
  curvature-aided gradient method for distributed optimization,'' \emph{2018
  IEEE Conference on Decision and Control (CDC)}, pp. 1751--1756, 2018.

\bibitem{Wai2017CurvatureaidedIA}
H.~Wai, W.~Shi, A.~Nedic, and A.~Scaglione, ``Curvature-aided incremental
  aggregated gradient method,'' \emph{2017 55th Annual Allerton Conference on
  Communication, Control, and Computing (Allerton)}, pp. 526--532, 2017.

\bibitem{Mao2020WalkmanAC}
X.~Mao, K.~Yuan, Y.~Hu, Y.~Gu, A.~H. Sayed, and W.~Yin, ``Walkman: A
  communication-efficient random-walk algorithm for decentralized
  optimization,'' \emph{IEEE Transactions on Signal Processing}, vol.~68, pp.
  2513--2528, 2020.

\bibitem{Boyd2006RandomizedGA}
S.~P. Boyd, A.~Ghosh, B.~Prabhakar, and D.~Shah, ``Randomized gossip
  algorithms,'' \emph{IEEE Transactions on Information Theory}, vol.~52, pp.
  2508--2530, 2006.

\bibitem{Koloskova2019DecentralizedSO}
A.~Koloskova, S.~Stich, and M.~Jaggi, ``Decentralized stochastic optimization
  and gossip algorithms with compressed communication,'' \emph{ArXiv}, vol.
  abs/1902.00340, 2019.

\bibitem{Aysal2009BroadcastGA}
T.~Aysal, M.~E. Yildiz, A.~Sarwate, and A.~Scaglione, ``Broadcast gossip
  algorithms for consensus,'' \emph{IEEE Transactions on Signal Processing},
  vol.~57, pp. 2748--2761, 2009.

\bibitem{Duchi2012DualAF}
J.~C. Duchi, A.~Agarwal, and M.~Wainwright, ``Dual averaging for distributed
  optimization: Convergence analysis and network scaling,'' \emph{IEEE
  Transactions on Automatic Control}, vol.~57, pp. 592--606, 2012.

\bibitem{duchi2013local}
J.~C. Duchi, M.~I. Jordan, and M.~J. Wainwright, ``Local privacy and
  statistical minimax rates,'' in \emph{2013 IEEE 54th Annual Symposium on
  Foundations of Computer Science}.\hskip 1em plus 0.5em minus 0.4em\relax
  IEEE, 2013, pp. 429--438.

\bibitem{kairouz2014extremal}
P.~Kairouz, S.~Oh, and P.~Viswanath, ``Extremal mechanisms for local
  differential privacy,'' in \emph{Advances in neural information processing
  systems}, 2014, pp. 2879--2887.

\bibitem{Wu2016DifferentiallyPS}
X.~Wu, A.~Kumar, K.~Chaudhuri, S.~Jha, and J.~F. Naughton, ``Differentially
  private stochastic gradient descent for in-rdbms analytics,'' \emph{ArXiv},
  vol. abs/1606.04722, 2016.

\bibitem{Bassily2014PrivateER}
R.~Bassily, A.~D. Smith, and A.~Thakurta, ``Private empirical risk
  minimization: Efficient algorithms and tight error bounds,'' \emph{2014 IEEE
  55th Annual Symposium on Foundations of Computer Science}, pp. 464--473,
  2014.

\bibitem{Gandikota2019vqSGDVQ}
V.~Gandikota, R.~K. Maity, and A.~Mazumdar, ``vqsgd: Vector quantized
  stochastic gradient descent,'' \emph{ArXiv}, vol. abs/1911.07971, 2019.

\bibitem{Xiong2016RandomizedRW}
S.~Xiong, A.~D. Sarwate, and N.~B. Mandayam, ``Randomized requantization with
  local differential privacy,'' \emph{2016 IEEE International Conference on
  Acoustics, Speech and Signal Processing (ICASSP)}, pp. 2189--2193, 2016.

\bibitem{Chen2020BreakingTC}
W.-N. Chen, P.~Kairouz, and A.~{\"O}zg{\"u}r, ``Breaking the
  communication-privacy-accuracy trilemma,'' \emph{ArXiv}, vol. abs/2007.11707,
  2020.

\bibitem{Zhao2014StochasticOW}
P.~Zhao and T.~Zhang, ``Stochastic optimization with importance sampling,''
  \emph{ArXiv}, vol. abs/1401.2753, 2014.

\bibitem{Ayache2019RandomWG}
G.~Ayache and S.~E. Rouayheb, ``Random walk gradient descent for decentralized
  learning on graphs,'' \emph{2019 IEEE International Parallel and Distributed
  Processing Symposium Workshops (IPDPSW)}, pp. 926--931, 2019.

\bibitem{Boyd2007SubgradientM}
D.~P. Bertsekas, ``Nonlinear programming,'' \emph{Journal of the Operational
  Research Society}, vol.~48, p. 334, 1995.

\bibitem{Dwork2014TheAF}
C.~Dwork and A.~Roth, ``The algorithmic foundations of differential privacy,''
  \emph{Foundations and Trends in Theoretical Computer Science}, vol.~9, pp.
  211--407, 2014.

\bibitem{balle}
M.~Bun and T.~Steinke, ``Concentrated differential privacy: Simplifications,
  extensions, and lower bounds,'' \emph{ArXiv}, vol. abs/1605.02065, 2016.

\bibitem{Levin}
D.~A. Levin, Y.~Peres, and E.~L. Wilmer, \emph{{Markov chains and mixing
  times}}.\hskip 1em plus 0.5em minus 0.4em\relax American Mathematical
  Society, 2006.

\bibitem{Robbins2007ASA}
H.~E. Robbins, ``A stochastic approximation method,'' \emph{Annals of
  Mathematical Statistics}, vol.~22, pp. 400--407, 2007.

\bibitem{Bottou2010LargeScaleML}
L.~Bottou, ``Large-scale machine learning with stochastic gradient descent,''
  in \emph{COMPSTAT}, 2010.

\bibitem{Bach2011NonAsymptoticAO}
F.~R. Bach and E.~Moulines, ``Non-asymptotic analysis of stochastic
  approximation algorithms for machine learning,'' in \emph{NIPS}, 2011.

\bibitem{lee}
C.-H. Lee, X.~Xu, and D.~Y. Eun, ``{Beyond random walk and metropolis-hastings
  samplers: why you should not backtrack for unbiased graph sampling},'' in
  \emph{SIGMETRICS}, 2012.

\bibitem{Chen2019FastAA}
B.~Chen, Y.~Xu, and A.~Shrivastava, ``Fast and accurate stochastic gradient
  estimation,'' in \emph{NeurIPS}, 2019.

\bibitem{Borsos2019OnlineVR}
Z.~Borsos, S.~Curi, K.~Y. Levy, and A.~Krause, ``Online variance reduction with
  mixtures,'' in \emph{ICML}, 2019.

\bibitem{Alain2015VarianceRI}
G.~Alain, A.~Lamb, C.~Sankar, A.~C. Courville, and Y.~Bengio, ``Variance
  reduction in sgd by distributed importance sampling,'' \emph{ArXiv}, vol.
  abs/1511.06481, 2015.

\bibitem{bouchard2015online}
G.~Bouchard, T.~Trouillon, J.~Perez, and A.~Gaidon, ``Online learning to
  sample,'' 2015.

\bibitem{Bottou2007TheTO}
L.~Bottou and O.~Bousquet, ``The tradeoffs of large scale learning,'' in
  \emph{NIPS}, 2007.

\bibitem{Bach2013NonstronglyconvexSS}
F.~R. Bach and E.~Moulines, ``Non-strongly-convex smooth stochastic
  approximation with convergence rate o(1/n),'' in \emph{NIPS}, 2013.

\bibitem{Duchi2011ErgodicMD}
J.~C. Duchi, A.~Agarwal, M.~Johansson, and M.~I. Jordan, ``Ergodic mirror
  descent,'' \emph{2011 49th Annual Allerton Conference on Communication,
  Control, and Computing (Allerton)}, pp. 701--706, 2011.

\bibitem{Shi2015EXTRAAE}
W.~Shi, Q.~Ling, G.~Wu, and W.~Yin, ``Extra: An exact first-order algorithm for
  decentralized consensus optimization,'' \emph{SIAM J. Optim.}, vol.~25, pp.
  944--966, 2015.

\bibitem{Chen2012FastDF}
A.~I. Chen, ``Fast distributed first-order methods,'' 2012.

\bibitem{Jakoveti2014FastDG}
D.~Jakoveti{\'c}, J.~Xavier, and J.~M.~F. Moura, ``Fast distributed gradient
  methods,'' \emph{IEEE Transactions on Automatic Control}, vol.~59, pp.
  1131--1146, 2014.

\bibitem{Koloskova2020AUT}
A.~Koloskova, N.~Loizou, S.~Boreiri, M.~Jaggi, and S.~Stich, ``A unified theory
  of decentralized sgd with changing topology and local updates,''
  \emph{ArXiv}, vol. abs/2003.10422, 2020.

\bibitem{Gammawiki}
A.~Papoulis and S.~Pillai, ``Probability, random variables, and stochastic
  processes, fourth edition,'' 2002.

\bibitem{aerinKim}
A.~Kim, ``Beta distribution: Intuition, examples, and derivation,'' January
  2020,
  {https://towardsdatascience.com/beta-distribution-intuition-examples-and-derivation-cf00f4db57af},
  {[Online; posted 08-January-2020]}.

\bibitem{ar}
\BIBentryALTinterwordspacing
``{Every Convex Function is Locally Lipschitz},'' \emph{The American
  Mathematical Monthly}, vol.~79, no.~10, pp. 1121--1124, 1972. [Online].
  Available: \url{http://www.jstor.org/stable/2317434}
\BIBentrySTDinterwordspacing

\bibitem{convex}
J.~C. Dattorro, ``Convex optimization and euclidean distance geometry,'' 2004.

\end{thebibliography}

\end{document}